\documentclass[preprint,aps,prc,nofootinbib,showpacs]{revtex4-1}

\usepackage{mathrsfs,amsmath,amssymb,amsfonts}
\usepackage{graphicx,graphics,epsfig,hhline,color}% Include figure files
\usepackage{natbib}

\newcommand{\be}{\begin{eqnarray}}
\newcommand{\ee}{\end{eqnarray}}

\newcommand{\Z}{\mathbb{Z}}
\newcommand{\lagrange}{{\cal L}}

\newcommand{\expm}{e^{-\beta\,E^+_p}}
\newcommand{\expmm}{e^{-2\beta\,E^+_p}}
\newcommand{\expmmm}{e^{-3\beta\,E^+_p}}
\newcommand{\expp}{e^{-\beta \,E^-_p}}
\newcommand{\expppp}{e^{-3\beta\,E^-_p}}
\newcommand{\exppp}{e^{-2\beta\,E^-_p}}

\newcommand{\bc}{\begin{center}}
\newcommand{\ec}{\end{center}}

\newcommand{\Tcc}{T^\chi_c}

\newcommand{\bea}{\begin{eqnarray}}
\newcommand{\eea}{\end{eqnarray}}

\newcommand{\beq}{\begin{eqnarray}}
\newcommand{\eeq}{\end{eqnarray}}
\newcommand{\Tr}{\mathrm{Tr}}

\newcommand{\pslash}{{p\hspace*{-0.14truecm}\slash}}

\begin{document}

\title{Phase diagram and critical properties within an effective model of QCD:
the Nambu--Jona-Lasinio model coupled to the Polyakov loop}
\author{\bf P. Costa}
\affiliation{Departamento de F\'{\i}sica, Universidade de
Coimbra, P-3004-516 Coimbra}
\affiliation{E.S.T.G., Instituto Polit\'ecnico de Leiria,
Morro do Lena-Alto do Vieiro, 2411-901 Leiria, Portugal}
\author{\bf M. C. Ruivo}
\author{\bf C. A. de Sousa}
\affiliation{Centro de F\'{\i}sica Computacional,
Departamento de F\'{\i}sica, Universidade de Coimbra, P-3004-516 Coimbra,
Portugal}
\author{\bf H. Hansen}
\affiliation{IPNL, Universit\'e de Lyon/Universit\'e Lyon 1, CNRS/IN2P3, 4 rue E.Fermi,
F-69622 Villeurbanne Cedex, France}

%
%\altaffiliation[Corresponding author]{}
%\email{pcosta@teor.fis.uc.pt}

%\email{}

\date{\today}
%

%======================================================================
%======================================================================
\begin{abstract}
We investigate the phase diagram of the so-called
Polyakov--Nambu--Jona-Lasinio model at finite temperature and
non-zero chemical potential with three quark flavors. Chiral and
deconfinement phase transitions are discussed and the relevant
order-like parameters are analyzed. The results are compared with
simple thermodynamic expectations and lattice data. We present the
phase diagram in the ($T,\,\mu_B)$ plane, paying special attention
to the critical end point: as the strength of the flavor-mixing
interaction becomes weaker, the critical end point moves to low
temperatures and can even disappear.
\end{abstract}
%======================================================================
%
\maketitle

%======================================================================
%======================================================================
\section{Introduction}

Symmetries play a fundamental role in physics. In fact, the modern
fundamental physics is dominated by considerations underlying
symmetries which can be exact, approximate (explicitly broken) and
(spontaneously) broken. A special role is played by gauge or local
symmetries which lead to the description of the ``real world'' as
the so-called local/gauge theories with spontaneously broken symmetries.

The concept of spontaneous symmetry breaking has been  transferred
from condensed matter physics to quantum field theory by  Nambu
\cite{Nambu:1960xd}. It has been introduced in particle physics on
the grounds of an analogy with the breaking of (electromagnetic)
gauge symmetry in the theory of superconductivity by Bardeen, Cooper
and  Schrieffer (the so-called BCS theory). The application of
spontaneous symmetry breaking to particle physics in the 1960s and
successive years led to profound physical consequences and played a
fundamental role in the edification of several models of elementary
particles.

Spontaneous breaking of chiral symmetry, in particular, is known to
govern the low-energy properties of hadrons \cite{Leutwyler:2001hn,
Weinberg:1978kz,Gasser:1984gg}.
Some QCD-like models have been proposed before the advent of QCD, and
the phenomenon of spontaneous breaking of chiral symmetry and
Nambu{--}Goldstone theorem was established more than 40 years ago
\cite{Nambu:1960xd}.

The Nambu--Jona-Lasinio (NJL) model was proposed in 1961 to explain
the origin of the nucleon mass with the help of spontaneous breaking
of chiral symmetry \cite{Nambu:1961tp,Nambu:1961fr}. At that time,
the model was formulated in terms of nucleons, pions and scalar
sigma mesons.
The introduction of the quark degrees of freedom and the description
of hadrons  by Eguchi and Kikkawa
\cite{Eguchi:1976iz,Kikkawa:1976fe} in the chiral limit, where the
bare quark mass is $m_0=0$, and a more realistic version with
$m_0\neq 0$ by Volkov and Ebert
\cite{Volkov:1982zx,Ebert:1982pk,Volkov:1984kq}, initiate a very
intensive activity by several research groups \cite{Volkov:2005kw}.

There is strong evidence that quantum chromodynamics (QCD) is the
fundamental theory of strong interactions. Its basic constituents
are quarks and gluons that are confined in hadronic matter. It is
believed that at high temperatures or densities the hadronic matter
should undergo a phase transition into a new state of matter, the
quark-gluon plasma (QGP). A challenge of theoretical studies based
on QCD is to predict the equation of state, the critical point and
the nature of the phase transition.

As the evolution of QCD at finite density/temperature is very
complicated, QCD-like models, as for instance NJL type models, have
been developed providing guidance and information relevant to
observable experimental signs of deconfinement and QGP features.
In fact, there has been great progress in the understanding  of the
properties of matter under extreme conditions of density and/or
temperature, where the restoration of symmetries (e.g., the chiral
symmetry) and the phenomenon of deconfinement should occur. These
extreme conditions might be achieved in ultrarelativistic heavy-ion
collisions or in the interior of neutron stars.
In this context, increasing attention has been devoted to the study
of the modification of particles propagating in a hot or dense
medium \cite{Mannarelli:2005pz,Kitazawa:2005mp}. The possible
survival of bound states in the deconfined phase of QCD
\cite{Datta:2003ww,Shuryak:2004tx,Shuryak:2004cy,Wong:2004zr,Wong:2006dz,Datta:2004wc,Alberico:2005xw,Mocsy:2005qw}
has also opened interesting scenarios for the identification of the
relevant degrees of freedom in the vicinity of the phase transition
\cite{Koch:2005vg,Ejiri:2005wq,Liao:2005pa}. Besides lattice
calculations
\cite{deForcrand:2001gv,Pushkina:2004wa,Petreczky:2003iz,Wissel:2005pb},
high temperature properties of QCD can be studied, starting from the
QCD Lagrangian, within different theoretical schemes, like the
dimensional reduction \cite{Hansson:1991kb,Laine:2003bd} or the hard
thermal loop approximation
\cite{Alberico:2004we,Alberico:2006wc,Karsch:2000gi}. Actually both
the above approaches rely on a separation of momentum scales which,
strictly speaking, holds only in the weak coupling regime $g\ll 1$.
Hence they cannot tell us anything about what happens in the
vicinity of the phase transition. On the other hand, a system close
to a phase transition is characterized by large correlation lengths
(infinite in the case of a second order phase transition). Its
behavior is mainly driven by the symmetries of the Lagrangian,
rather than by the details of the microscopic interactions.

Confinement and chiral symmetry breaking are two of the most
important features of QCD. As already referred, chiral models like
the NJL  model
\cite{Nambu:1961tp,Nambu:1961fr,Klevansky:1992qe,Buballa:2003qv}
have been successful in explaining the dynamics of spontaneous
breaking of chiral symmetry and its restoration at high temperatures
and densities/chemical potentials.  Recently, this and other type of
models, together with an intense experimental activity, are underway
to construct the phase diagram of QCD.

The actual NJL type models describe interactions between
constituent quarks, giving the correct chiral properties, and offers
a simple and practical illustration of the basic mechanisms that
drive the spontaneous breaking of chiral symmetry, a key feature of
QCD in its low temperature and density phase. In order to take into
account features of both chiral symmetry breaking and deconfinement,
static degrees of freedom are introduced in this Lagrangian through
an effective gluon potential in terms of the Polyakov loop
\cite{Meisinger:1995ih,Pisarski:2000eq,Pisarski:2002ji,Meisinger:2001cq,Fukushima:2003fw,Mocsy:2003qw,Ratti:2005jh,Hansen:2006ee}.
The coupling of the quarks to the Polyakov loop leads to the
reduction of the weight of the quark degrees of freedom at low
temperature as a consequence of the restoration of the $\Z_{N_c}$
symmetry associated with the color confinement.

In first approximation, the behavior of a system ruled by QCD is
governed by the symmetry properties of the Lagrangian, namely the
(approximate) global symmetry
$\mbox{SU}_L(N_f)\times\mbox{SU}_R(N_f)$, which is spontaneously
broken to SU$_V(N_f)$, and the (exact) SU$_c(N_c)$ local color
symmetry.
Indeed, in NJL type models the mass of a constituent quark is
directly related to the chiral condensate, which is the order
parameter of the chiral phase transition and, hence, is
non-vanishing at zero temperature and density. Here the system lives
in the phase of spontaneously broken chiral symmetry: the strong
interaction, by polarizing the vacuum and turning it into a
condensate of quark-antiquark pairs, transforms an initially
point-like quark with its small bare mass $m_0$ into a massive
quasiparticle with a finite size.
Despite their widespread use, NJL models suffer a major shortcoming:
the reduction to global (rather than local) color symmetry prevents
quark confinement.

On the other hand, in a non-abelian pure gauge theory, the Polyakov
loop serves as an order parameter for the transition from the low
temperature, $\mathbb{Z}_{N_c}$ symmetric, confined phase (the
active degrees of freedom being color-singlet states, the
glueballs), to the high temperature, deconfined phase (the active
degrees of freedom being colored gluons), characterized by the
spontaneous breaking of the $\mathbb{Z}_{N_c}$ (center of
SU$_c(N_c)$) symmetry.
With the introduction of dynamical quarks, this symmetry breaking
pattern is no longer exact: nevertheless it is still possible to
distinguish a hadronic (confined) phase from a QGP (deconfined) one.

In the PNJL model, quarks are coupled simultaneously to the chiral
condensate and to the Polyakov loop: the model includes features of
both chiral and $\Z_{N_c}$ symmetry breaking. The model has proven
to be successful in reproducing lattice data concerning QCD
thermodynamics \cite{Ratti:2005jh}. The coupling to the Polyakov
loop, resulting in a suppression of the unwanted quark contributions
to the thermodynamics below the critical temperature, plays a
fundamental role for the analysis of the critical behavior.

One of the important features of the QCD phase diagram is the
existence of a phase boundary in the $(T,\,\mu_B)$ that separates
the chirally broken hadronic phase from the chirally symmetric QGP
phase.
Arguments based on effective model calculations  suggest that the
QCD phase diagram can exhibit a tricritical point (TCP)/critical end
point (CEP) where the line of first order matches that of second
order/analytical crossover
\cite{Asakawa:1989bq,Barducci:1989wi,Barducci:1989eu,Barducci:1993bh,Stephanov:2004wx,Casalbuoni:2006rs}.
The discussion about the existence and location of  such critical
points  of QCD  is a very important topic nowadays \cite{CEP}.

%........................................................................................

This paper is organized as follows.
In Section \ref{QCD} we analyze the main features and symmetries of QCD.
In Section \ref{formalism} we present the model and formalism starting with the deduction of the self consistent
equations. We also obtain  the equations of state and  the response functions. The regularization
procedure used in the model calculations is also included.
Section \ref{P_T_finite_T} is devoted to  the study of the equation of state at finite temperature.
In Section \ref{phase} we study the phase diagram and the location of the critical end point.
In Section \ref{sec_isent} we discuss the important role of the choice of the model parameters for the correct
description of isentropic trajectories.
In Section \ref{CEP1} we analyze the effects of strangeness and anomaly strength on the location of the critical
end point.
In Section \ref{CEP2} we proceed to study the size of the critical region around the critical end point and
its consequences for the susceptibilities.
Finally, concluding remarks are presented in Section \ref{conclusions}.

%======================================================================
%======================================================================

%======================================================================
\section{Symmetries of QCD}\label{QCD}

%======================================================================
\subsection{Quantum Chromodynamics}

The Lagrangian of QCD is written as \cite{Weinberg:1973un,Fritzsch:1973pi}
%%%%%%%
\begin{equation}
\mathcal{L}_{QCD}\;=\;\bar{q}\,(\,i\gamma^{\mu}D_{\mu}-\hat{m}\,)\,q\;-\;\frac
{1}{4}F_{\mu\nu}^{a\,}\,F_{a}^{\mu\nu} \label{Lqcd}%
\end{equation}
%%%%%%%
where $q$ is the quark field with six flavors ($u,d,s,c,b,t$), three colors ($N_{c}=3$)
and $\hat{m}$ being the corresponding current quark mass matrix in flavor space
($\hat{m}=\mathrm{diag}_{f}(m_{u},m_{d},\dots)$). The covariant derivative,
%%%%%%%
\begin{equation}
D_{\mu}\;=\;\partial_{\mu}\;-\;igt^{a}\,A_{\mu}^{a}%
\end{equation}
%%%%%%%
incorporates the color gauge field $A_{\mu}^{a}$ ($a=1,2,...,8$) and
%%%%%%%
\begin{equation}
F_{\;\mu\nu}^{a}\;=\;\partial_{\mu}\,A_{\nu}^{a}\;-\;\partial_{\nu}\,A_{\mu
}^{a}\;+\;g\,f^{abc}\,A_{\mu}^{b}\,A_{\nu}^{c} \label{Gamn}%
\end{equation}
%%%%%%%
is the gluon field strength tensor; $t^{a}$ is the Gell-Mann color matrix in SU(3)
\linebreak ($[t^{a},t^{b}]=if_{abc}t^{c},\mathrm{tr}(t^{a}t^{b})=\frac{\delta^{ab}}{2}$) and
$f^{abc}$ are the corresponding antisymmetric structure constants. Finally, $g$ is the
QCD coupling constant.

The QCD Lagrangian is by construction symmetric under SU(3) gauge
transformations in color space and because of the non-Abelian
character of the gauge group QCD has some main features: it is a
\textit{renormalizable} quantum field theory \cite{'tHooft:1971fh}
with a single coupling constant for both, the quark--gluon
interactions and the gluonic self--couplings involving vertices with
three and four gluons; it has \textit{confinement}, {\em i.e.},
objects carrying color like quarks and gluons do not exist as
physical degrees of freedom in the vacuum. In addition,  QCD is a
theory that has \textit{asymptotic freedom}
\cite{Gross:1973id,Politzer:1973fx}, {\em i.e.}, for large momenta,
$Q$, or wavelengths of the order of $10^{-1}$ fm (ultraviolet
region) the couplings are weak and the quarks and gluons propagate
almost freely.  For low momenta, or wavelengths of about 1 fm
(infrared region), it happens the opposite situation and the
couplings are quite strong. The system is now highly non
perturbative. According to this property, the attraction between two
quarks grows indefinitely as they move away from each other. This
implies that the interaction between quarks and gluons can not be
treated perturbatively, making the perturbative treatment of QCD
not applicable to describe hadrons with masses below $\sim$2 GeV.

The low energy regime is specially interesting once it is relevant
to the study of hadronic properties as for example in low energy QCD
and nuclear physics.

The non-perturbative structure of the vacuum is characterized by the
existence of quark condensates, {\em i.e.}, it is expected non-zero
values for the scalar density $\left\langle \bar{q}q\right\rangle$,
by the appearance of light pseudoscalar particles, who are
identified with (quasi) Goldstone bosons
\cite{Goldstone:1961eq,Goldstone:1962es}, and also by the existence
of gluon pairs \cite{Hatsuda:1994pi}.

On the other hand, if QCD has a mechanism that includes the
confinement of quarks the mass parameters $m_i$ are not observable
quantities. However, they can be estimated in terms of the masses of
some hadronic observables through current algebra's methods. These
masses are denominated by \textit{current} quark masses to
distinguish them from the \textit{constituent} quark masses, which
are effective masses generated by the spontaneous breaking of chiral
symmetry in phenomenological models of quarks.

%======================================================================
\subsection{Chiral Symmetry Breaking} \label{subSec:Chiral symmetry}

Due to the relevant role that spontaneous breaking of chiral
symmetry plays in hadronic physics at low energies, this symmetry is
one of most important symmetries of QCD. Here we will concentrate in
the $N_f=3$ case. In the chiral limit, {\em i.e.}
$m_{u}=m_{d}=m_{s}=0$, QCD is chiral invariant which means that the
QCD Lagrangian (\ref{Lqcd}) is invariant under the group of
symmetries:
%%%%%
\begin{equation}
\text{U(3)}_{L}\otimes\text{U(3)}_{R}\text{=SU}_{V}\text{(3)}\otimes
\text{SU}_{A}\text{(3)}\otimes\text{U}_{V}\text{(1)}\otimes\text{U}_{A}\text{(1)}.
\end{equation}
%%%%%

These symmetries are presented in Table \ref{table:simetrias} where
it is also possible to see the transformations under which the
Lagrangian is invariant, the currents that are conserved according
to Noether's theorem and the respective manifestations  of the
symmetries in nature.

%%%%%%%%%%%%%%%%%%%%%%%%%%%%%%%%%%%
\begin{table}[t]
\begin{center}
{\small
\begin{tabular}[c]{|c|c|c|c|c|}\hline
{\bf Symmetry} & {\bf Transformation} & {\bf Current} & {\bf Name} & {\bf Manifestation in nature}
\\\hline\hline
SU$_{V}$(3) & $q\rightarrow\exp(-i\frac{\lambda_{a}\alpha_{a}}{2})q$ &
$V_{\mu}^{a}=\bar{q}\gamma_{\mu}\frac{\lambda_{a}}{2}q$ & Isospin & Approximately
conserved
\\
U$_{V}$(1) & $q\rightarrow\exp(-i\alpha_{V})q$ & $V_{\mu}=\bar{q}\gamma_{\mu}q$ & Baryonic
& Conserved
\\
SU$_{A}$(3) & $q\rightarrow\exp(-i\frac{\gamma_{5}\lambda_{a}\theta_{a}}{2})q$ &
$A_{\mu}^{a}=\bar{q}\gamma_{\mu}\gamma_{5}\frac{\lambda_{a}}{2}q$ & Quiral &
Spontaneously broken
\\
U$_{A}$(1) & $q\rightarrow\exp(-i\gamma_{5}\alpha_{A})q$ & $A_{\mu}=\bar{q}\gamma_{\mu}\gamma_{5}q$ & Axial & `` U$_{A}$(1) problem''%
\\\hline
\end{tabular}
\caption{\label{table:simetrias} QCD symmetries in the chiral limit.}%
}

\end{center}
\end{table}
%%%%%%%%%%%%%%%%%%%%%%%%%%%%%%%%%%%

The SU$_{V}$(3) and U$_{V}$(1)  symmetries ensure  the conservation
of isospin and baryon number, respectively, while   the SU$_{A}$(3)
and U$_{A}$(1) symmetries are transformations that involve the
$\gamma_{5}$ matrix and therefore alter the parity of the state in
which they operate. For the sake of uniformity throughout the text
we will designate by chiral symmetry the SU$_{A}$(3) symmetry and by
axial symmetry the U$_{A}$(1) symmetry.

From the experimental point of view the manifestation of chiral
symmetry would be the existence of parity doublets, {\em i.e.}, a
multiplet of particles with the same mass and opposite parity for
each multiplet of isospin (the chiral partners), in the hadronic
spectrum; that situation is not verified.
Similarly, if the symmetry U$_{A}$(1) would manifest itself,
the existence of a partner with opposite parity to each hadron
should be observed experimentally. As neither of these situations is
observed in the hadron spectrum, these symmetries must be somehow
broken.

Concerning the SU$_{A}$(3) symmetry, the theory must contain a
mechanism for the \textit{spontaneous breaking of chiral symmetry},
which represents a transition to an asymmetric phase. This is
closely related to the existence of non-zero quark condensates,
$\left\langle\bar{q}q\right\rangle$, which are not invariant under
SU$_{A}$(3) transformations and therefore act as order parameters
for the spontaneously broken chiral symmetry.

According to the Goldstone theorem, the spontaneous breaking of a
continuous global symmetry implies the existence of a particle with
zero mass, the Goldstone boson. In the case we are considering, the
symmetry breaking is closely related to the appearance of eight
degenerate Goldstone bosons with zero mass. As matter of fact the
pions where the first mesons associated to Goldstone bosons due to
their small mass. Indeed, if compared to the mass of the nucleon one
has $M_{\pi}/M_{N}=0.15$. To reproduce the meson spectrum it is also
necessary that the theory incorporates a mechanism which explicitly
breaks the chiral symmetry: the Lagrangian must include perturbative
terms that break \textit{ab initio} the symmetry allowing the
lifting of the degeneracy in the pseudoscalar mesons spectrum. These
terms are the current quark masses $m_{u}\,m_{d}$ and $m_{s}$.

Nevertheless, the fundamentals of this process are not yet
completely understood at the level of QCD. However, the chiral
symmetry breaking, the generation of the constituent quark masses
and the concomitant appearance of Goldstone bosons are explained in
a very satisfactory way in certain theories and physical models like
NJL type models.

Another symmetry effect occurs in the $\eta-\eta'$ sector. The
pseudoscalar $\eta'$ meson is not of the Goldstone type.
Classically, the QCD Lagrangian with $N_f=3$ massless flavors
possesses the U$_L(3)\times$U$_R(3)$ symmetry. In fact, the
Adler--Jackiw--Bell U$_A(1)$ anomaly breaks this symmetry to
SU$_L(3)$$\times$ SU$_R(3)$ and yields a large mass for the $\eta'$
even if quark masses are zero. The origin of this anomaly are the
instantons in QCD as explained in the next subsection.

In conclusion, QCD with three massless quarks possesses the
SU$_L$(3)$\times$ SU$_R$(3) global chiral symmetry. At low energy,
the pertinent degrees of freedom are mesons indicating that QCD
experiences a confined/deconfined phase transition when energy
increases. The study of mesons give us insight on the
non-perturbative vacuum of QCD at low energy.

%======================================================================
\subsection{U$_A$(1) Symmetry Breaking} \label{subSec:UA(1)}

Now we will analyze the U$_{A}$(1) symmetry. It is known that in the
chiral limit, and at the classical level, $\mathcal{L}_{QCD}$ is
also invariant under the axial  transformation (see Table \ref{table:simetrias}).

As already mentioned, the absence in nature of chiral partners with
the same mass, related with the U$_{A}$(1) symmetry, opens the
possibility that U$_{A}$(1) symmetry is also spontaneously broken,
similarly to what happens with the SU$_{A}$(3) symmetry.
Consequently, there should exist another pseudoscalar Goldstone
boson. S. Weinberg estimated the mass of this particle, outside the
chiral limit, at about $\sqrt{3}M_{\pi}$ \cite{Weinberg:1975ui}.

From all hadrons known in nature, the only candidates with the correct quantum numbers
were the $\eta(549)$ and $\eta^{\prime}(985)$. However, both violate the Weinberg
estimate because their masses are too high: the $\eta^{\prime}(985)$ has a mass
comparable to the mass of nucleons (``$\eta^{\prime}$ problem'') and, furthermore, the
$\eta(549)$ is identified as belonging to the octet of pseudoscalar mesons. 
If a particle with the characteristics pointed out by Weinberg does
not exist, then where is the ninth Goldstone boson? 
If this Goldstone boson does not exist there is no spontaneous
breaking of U$_{A}$(1) symmetry.

In $1976$ G. 't Hooft suggested that the U$_A$(1) symmetry does not
exist at the quantum level, being explicitly broken by the axial
anomaly \cite{Weinberg:1975ui} which can be described at the
semiclassical level by instantons
\cite{'tHooft:1976up,'tHooft:1976fv}.
Instantons are $4D$ topologically stable soliton solutions of QCD.
Other well known solitons are $2D$ Abrikosov vortices in
superconducting material or $3D$ magnetic monopole in QCD. The
instantons ``transform'' left handed fermions into right handed ones
(and conversely). The transition induced by this solution of QCD has
a non zero axial charge variation: $\Delta Q_5 = \pm 2 N_f$, {\em
i.e.}, one left handed fermion ($+1$ in axial charge) transforms in
a right handed fermion ($-1$). To mimic this interaction in a purely
fermionic effective theory, 't Hooft proposed the following
interaction Lagrangian, the so-called ``'t Hooft determinant'':
%%%%%%%
\bea \lagrange_{inst} = \kappa e^{\imath \theta_{inst}} \det_{flavor} (\bar\psi_R(x)
\psi_L(x)) + h.c. \eea
%%%%%%%
This interaction ``absorbs'' $N_f$ left helicity fermions and
converts them to right handed ones (and conversely).  In the
following, we will take $\theta_{inst} = 0$.

In this context, as suggested by 't Hooft, the instantons can play a crucial role in
breaking explicitly the U$_{A}$(1) symmetry giving to the $\eta^{\prime}$ a mass of about
$1$ GeV outside the chiral limit.
This implies that the mass of $\eta^{\prime}$ has a different origin
than the other masses of the pseudoscalar mesons and can not be seen
as the missing Goldstone  boson due to spontaneous breaking of
U$_{A}$(1) symmetry. Consequently,  the U$_A$(1) anomaly is very
important once it is responsible for the flavor mixing effect that
removes the degeneracy among several mesons.

Finally, an interesting open question also related to the symmetry
that we wish to address is whether both chiral
SU$(N_f)\times$SU$(N_f)$ and axial U$_A$(1) symmetries are restored
in the high temperature/density phase and which observables could
carry information about these restorations.

%======================================================================
\subsection{The Polyakov Loop and the ${\mathbb Z}_3$ Symmetry Breaking: Pure Gauge Sector}

In this Section, following the arguments given in
\cite{Pisarski:2000eq,Pisarski:2002ji}, we discuss how the
deconfinement phase transition in a pure SU$(N_c)$ gauge theory can
be conveniently described through the introduction of an effective
potential for the complex Polyakov loop field, which we define in
the following.

Since we want to study the SU$(N_c)$ phase structure, first of all
an appropriate order parameter has to be defined. For this purpose
the Polyakov line
%%%%%%%
\be L \left(\vec{x}\right)&\equiv&\mathcal{P}\exp\left[i\int_{0}^{\beta}d\tau\,
A_4\left(\vec{x},\tau\right)\right]\label{eq:defpoly}
\ee
%%%%%%%
is introduced. In the above, $A_4 = i A_0$ is the temporal component of the Euclidean
gauge field $(\vec{A}, A_4)$, in which the strong coupling constant $g$ has been
absorbed, $\mathcal{P}$ denotes path ordering and the usual notation $\beta = 1/T$ has
been introduced with the Boltzmann constant set to one ($k_B = 1$).

The  Polyakov line $L\left(\vec{x}\right)$ can be described as an
operator of parallel transport of the gauge field
$A_4\left(\vec{x},\tau\right)$ into the direction $\tau$. One way to
understand why this quantity is indeed a parameter that can
distinguish between confined or deconfined phase is to consider the
two extremal behavior of $L$: a color field $A_4$ at the point
$(\vec x, \tau)$ will be transformed into the color field $L \times
A_4$ after being transported into the direction $\tau$. If
$L\rightarrow 1$ it means that nothing affects the propagation of
the field: the medium is in its deconfined phase. At the contrary
$L\rightarrow 0$ indicates that the color field cannot propagate in
the medium: it is confined.

Another way to see this point is to consider the variation of free
energy when an infinitely massive (hence static) quark (that acts as
a test color charge) is added to the system. To this purpose let us
introduce the Polyakov loop. When the theory is regularized on the
lattice, it reads:
%%%%%%%
\beq l(\vec{x})=\frac{1}{N_c} \Tr {L(\vec{x})}  \eeq
%%%%%%%
and it is a color singlet under SU$(N_c)$, but transforms
non-trivially, like a field of charge one, under $\Z_{N_c}$. Its
thermal expectation value can then be chosen as an order parameter
for the deconfinement phase transition
\cite{Polyakov:1978vu,'tHooft:1977hy,Svetitsky:1982gs}: in the usual
physical interpretation \cite{McLerran:1981pb,Rothe:2005nw},
$\langle l(\vec{x})\rangle$ is related to the change of free energy
occurring when a heavy color source in the fundamental
representation is added to the system. One has:
%%%%%%%
\beq \langle l(\vec{x})\rangle=e^{-\beta\Delta F_Q(\vec{x})} \label{criterion} \eeq
%%%%%%%
In the $\Z_{N_c}$ symmetric phase, $\langle l(\vec{x})\rangle=0$, implying that an
infinite amount of free energy is required to add an isolated heavy quark to the system:
in this phase color is confined.

In the case of the SU$(3)$ gauge theory, the Polyakov line
$L(\vec{x})$ gets replaced by its gauge covariant average over a
finite region of space, denoted as $\langle\!\langle
L(\vec{x})\rangle\!\rangle$ \cite{Pisarski:2000eq,Pisarski:2002ji}.
Note that $\langle\!\langle L(\vec{x})\rangle\!\rangle$ in general
is not a SU$(N_c)$ matrix.
The Polyakov loop field,
%%%%%%%
\beq \Phi(\vec x)\equiv\langle\!\langle l(\vec{x})\rangle\!\rangle=\frac 1 {N_c}\,\Tr_c
\, \langle\!\langle L(\vec x)\rangle\!\rangle \eeq
%%%%%%%
is then introduced.

Following the Landau-Ginzburg approach, a $\Z_3$ symmetric effective
potential is defined for the (complex) $\Phi$ field, which is
conveniently chosen to reproduce, at the mean field level, results
obtained in lattice
calculations \cite{Pisarski:2000eq,Pisarski:2002ji,Ratti:2005jh}. In
this approximation one simply sets the Polyakov loop field
$\Phi(\vec{x})$ equal to its expectation value $\Phi=const.$, which
minimizes the potential.

Concerning the effective potential for the (complex) field $\Phi$,
different choices are available in the literature
\cite{Ratti:2006gh,Roessner:2006xn,Fukushima:2008wg}; the one
proposed in \cite{Roessner:2006xn} (see Equation (\ref{Ueff})) is
known to give sensible results \cite{Roessner:2006xn,Sasaki:2006ww,Sasaki:2006ws} and will be
adopted in our parametrization of the PNJL model that will be
presented in Section \ref{formalism}. In particular, this
potential reproduces, at the mean field level, results obtained in
lattice calculations as it will be shown. The potential reads:
%%%%%%
\begin{equation}
    \frac{\mathcal{U}\left(\Phi,\bar\Phi;T\right)}{T^4}
    =-\frac{a\left(T\right)}{2}\bar\Phi \Phi +
    b(T)\mbox{ln}[1-6\bar\Phi \Phi+4(\bar\Phi^3+ \Phi^3)-3(\bar\Phi \Phi)^2]
    \label{Ueff}
\end{equation}
%%%%%%
where
%%%%%%
\begin{equation}
    a\left(T\right)=a_0+a_1\left(\frac{T_0}{T}\right)+a_2\left(\frac{T_0}{T}
  \right)^2\,\mbox{ and }\,\,b(T)=b_3\left(\frac{T_0}{T}\right)^3
\end{equation}
%%%%%

The effective potential exhibits the feature of a phase transition
from color confinement ($T<T_0$, { the minimum of the effective
potential being at $\Phi=0$}) to color deconfinement ($T>T_0$, the
minima of the effective potential occurring at $\Phi \neq 0$).

The parameters of the effective potential $\mathcal{U}$ are given in
Table \ref{table:param}. These parameters have been fixed in order
to reproduce the lattice data for the expectation value of the
Polyakov loop and QCD thermodynamics in the pure gauge sector
\cite{Kaczmarek:2002mc,Kaczmarek:2007pb}.

%%%%%%%%%%%%%%%%%%%%%%%%%
\begin{table}[t]
    \begin{center}
        \begin{tabular}{|c|c|c|c|}
            \hhline{|----|}
            $a_0$ & $a_1$ & $a_2$ & $b_3$ \\
            \hline\hline
            3.51  & -2.47 & 15.2 & -1.75  \\
            \hhline{|----|}
        \end{tabular}
         \caption{\label{table:param} Parameters for the effective potential in the pure gauge sector
         (Equation (\ref{Ueff})).}
    \end{center}
\end{table}
%%%%%%%%%%%%%%%%%%%%%%%%

The parameter $T_0$ is  the critical temperature for the
deconfinement phase transition within a pure gauge approach: it was
fixed to $270$ MeV, according to lattice findings. Different
criteria for fixing $T_0$  may be found in the literature, like in
\cite{Schaefer:2007pw}, where an explicit $N_f$ dependence of $T_0$
is presented by using renormalization group arguments.  However, it
is noticed that the Polyakov loop computed on the lattice with (2+1)
flavors and with fairly realistic quark masses is very similar to
the SU$_f$(2) case \cite{Kaczmarek:2007pb}.  Hence, it was chosen to
keep for the effective potential
$\mathcal{U}\left(\Phi,\bar\Phi;T\right)$ the same parameters which
were used in SU$_f$(2) PNJL~\cite{Roessner:2006xn}, including
$T_0=270$ MeV. The latter choice ensures an almost exact coincidence
between chiral crossover and deconfinement at zero chemical
potential, as observed in lattice calculations.
We notice, however, as it will be seen in Section
\ref{P_T_finite_T}, that a rescaling of $T_0$ to $210$ MeV may be
needed in some cases in order to get agreement between model
calculations and thermodynamic quantities obtained on the lattice.
Let us stress that this modification of $T_0$ (the only free
parameter of the Polyakov loop once the effective potential is
fixed) is essentially done for rescaling reasons (an absolute
temperature scale has not a very strong meaning in these kind of
Ginzburg--Landau model based on symmetry) but does not change
drastically the physics.

With the present choice of parameters, $\Phi$ and $\bar\Phi$ are
always lower than $1$ in the pure gauge sector. In any case, in the
range of applicability of our model, $T\leq (2-3)T_c$, there is a
good agreement between the results and the lattice data for $\Phi$.

To conclude this section, some comments are in order on the number
of parameters and the nature of the Polyakov field.

It should be noticed that the NJL parameters and the Polyakov
potential ones are not on the same footing. Whereas the NJL
parameters are directly related in a one to one correspondence with
a physical quantity, the Polyakov loop potential is there to insure
that the pure gauge lattice expectations are recovered. 
Hence the potential for the loop can be viewed as a
unique but functional parameter. The details of this function are
not very important to study the thermodynamic as soon as the
potential reproduces the lattice results. In order to clarify this
point, we remember that in a previous calculations in the $N_f = 2$
case  \cite{Costa:2009ae} we used two kind of potentials and
obtained very small differences in what concerns thermodynamics;
however, if one calculates susceptibilities with respect to $\Phi$
the log potential used here has to be preferred in order not to
have unphysical results.
The only true parameter is the pure gauge critical temperature $T_0$
that fixes the temperature scale of the system.  However we would
like to point out that it is expected in the Landau-Ginzburg
framework that a characteristic temperature for a phase transition
will not be a prediction: one needs to fix the correct energy scale
somehow and it is the role of this parameter. Hence we will allow
ourselves to  change this parameter in the calculation of some
observables in order to compare our results with lattice QCD
expectations.

Finally we want to stress that, contrarily to the full
Landau-Ginzburg effective field approach, the Polyakov loop
effective field is not a dynamical degree of freedom due to the the
lack of dynamical term in the Polyakov loop potential. Hence it is a
background gauge field in which quarks will propagate. Anyway the
potential mimics a pressure term that has the correct magnitude and
temperature behavior in order to get the Stefan-Boltzman limit for
the gluonic degrees of freedom and that explains the success of the
model.

%======================================================================
\subsection{QCD Phase Diagram}

Understanding the QCD phase structure is one of the most important
topics in the physics of strong interactions. The developed effort
on both the theoretical point of view---using effective models and
lattice calculations---and the experimental point of view is one of
the main goals of the heavy ion collisions program; it has proved
very fruitful, shedding light on properties of matter at finite
temperatures and densities.

From the theoretical point of view, the physics of the transition
from hadronic matter to the QGP is well established for baryonic
chemical potential $\mu_{B}=0$. Latest lattice results for (2+1)
flavors show a crossover transition at $T_c = (170 - 200)$ MeV
(using $m_q = 0.05\, m_s$) \cite{Cheng:2009zi}. Near the critical
temperature, $T_c$, the energy density (and other thermodynamic
quantities) show a strong growth, signaling the transition from a
hadronic resonance gas to a matter of deconfined quarks and gluons.
As matter of fact, the rapid rise of the energy density is usually
interpreted to be due to deconfinement {\em i.e.}, liberation of
many new degrees of freedom. In the limit where quark masses are
infinitely heavy the order parameter for the deconfinement phase
transition is the Polyakov loop. A rapid change in this quantity is
also an indication for deconfinement even in the presence of light
quarks \cite{Cheng:2009zi}.

Concerning chiral symmetry, and considering the chiral limit, it is
expected a chiral transition with the corresponding order parameter
being the quark condensate: the quark condensate vanishes at the
critical temperature $T_c$ and a genuine phase transition takes
place. Even away from the chiral limit, when the quark masses are
finite, it is expected, in the transition region, a ``crossover''
where the quark condensates rapidly drop indicating a partial
restoration of the chiral symmetry \cite{Cheng:2009zi}.

This rises the interesting question  whether the restoration of
chiral symmetry and the transition to the QGP occurs at same time.
In \cite{Adams:2005dq} it was proposed that the evidence of
restoration of chiral symmetry is a sufficient condition to
demonstrate the existence of a new state of matter, but it is not a
necessary condition for the discovery of the QGP. In fact, as
already pointed out, most of lattice results shows a tendency for
the restoration of chiral symmetry to happen simultaneously with
deconfinement, but this issue is not definitively determined from a
theoretical point of view.

At finite temperature and chemical potential the most common
three-flavor phase diagram shows a first order boundary of the
chiral phase transition separating the hadronic and quark phases.
This first--order line starts at non-zero chemical potential and
zero temperature and will finish in a point, the critical end point
($T^{CEP},\,\mu_B^{CEP}$), where the phase transition is of second
order. As the temperature increases and the chemical potential
decreases, the QCD phase transition becomes a crossover.

Although recent lattice QCD results by de Forcrand and Philipsen
question the existence of the CEP
\cite{deForcrand:2006pv,deForcrand:2007rq,deForcrand:2008vr}, this
critical point of QCD, proposed at the end of the eighties
\cite{Asakawa:1989bq,Barducci:1989wi,Barducci:1989eu,Barducci:1993bh},
is still a very important subject of discussion nowadays \cite{CEP}.
The search of the QCD CEP is one of the main goals in ``Super Proton
Synchrotron'' (SPS) at CERN \cite{Anticic:2009pe} and in the next
phase of the ``Relativistic Heavy Ion Collider'' (RHIC) running at
BNL.

When matter at high density and low temperature is considered, it is
expected that this type of matter is a color superconductor where
pairs of quarks condense (``diquark condensate''): this color
superconductor is a degenerate Fermi gas of quarks with a condensate
of Cooper pairs near the Fermi surface that induces color  Meissner
effects. With the color superconductivity it was realized that the
consideration of diquark condensates opened the possibility for a
wealth of new phases in the QCD phase diagram. At the  highest
densities/chemical potentials, where  the  QCD coupling is weak, the
ground state is a particularly symmetric state, the color-flavor
locked (CFL) phase: up, down, and strange quarks are paired in a
so-called color-flavor locked condensate. At lower densities the CFL
phase may be disfavored in comparison  with alternative new phases
which  may  break  translation  and/or  rotation invariance
\cite{Buballa:2003qv,Alford:2007xm}.

Recently it was argued that some features of hadron production in
relativistic nuclear collisions, mainly at SPS--CERN energies, can
be explained by the existence of a new form of matter beyond the
hadronic matter and the QGP: the so-called quarkyonic matter
\cite{McLerran:2007qj,Hidaka:2008yy,McLerran:2008ua,Fukushima:2008wg}.
It was also suggested that these different types of matter meet at a
triple point in the QCD phase diagram: both the hadronic matter, the
QGP, and the quarkyonic matter all coexist \cite{Andronic:2009gj}.

From the experimental point of view, different parts of the QCD
phase diagram have been investigated for different beam energies.
For high energies, corresponding to the energies of RHIC--BNL and
LHC--CERN (``Large Hadron Collider''), the matter is  produced at
high temperature and low baryonic density
\cite{Adams:2005dq,Adcox:2004mh,BraunMunzinger:2007zz}. On the other
hand, intermediate energy, corresponding to the energies of
``Alternating Gradient Synchrotron'' (AGS) in BNL or in SPS--CERN
\cite{Anticic:2009pe}, can be studied through matter formed at
moderate temperature and baryonic density. Finally, the energies
achieved in the GSI Heavy Ion Synchrotron (SIS) \cite {Hohne:2005qm}
and ``Nuclotron-based Ion Collider fAcility'' (NICA) at JINR \cite
{Sissakian:2010zz} will allow to obtain matter at low temperatures
and high baryonic densities. For example the main goal of the
research program on nucleus-nucleus collisions at GSI is the
investigation of highly compressed nuclear matter. Matter in this
form exists in neutron stars and in the core of supernova
explosions.

\begin{figure}[t]
    \begin{center}
    \vspace*{-0.5cm}
        \includegraphics[width=0.65\textwidth]{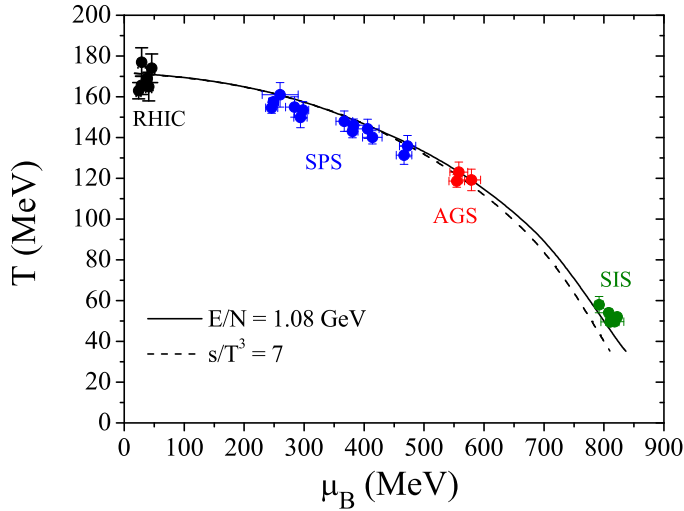}
    \end{center}
    \vspace*{-0.5cm}
\caption{The phase diagram from the experimental point of view: freeze-out points for the
different beam energies are showed \cite{Cleymans:2005xv}.}
\label{Fig:diag_phases}
\end{figure}

In Figure \ref{Fig:diag_phases} are indicated the points where the
hadrons cease to interact with each others ({\it freeze-out} points)
for the different beam energies. These points are related to final
state of the expansion of the \textit{fireball}.

It should also be noted that atomic nuclei by themselves represent a
system at finite density and zero temperature. At normal nuclear
density it is estimated that the quark condensate undergoes a
reduction of $30\%$ \cite{Brockmann:1996iv} so that the
resulting effects in the chiral order parameter can be measured in
beams of hadrons, electrons and photons in nuclear targets.

If the evidence for deconfinement can be found experimentally, then
the search for manifestations of chiral symmetry restoration will
be, in the near future, one of the main goals once the investigation
of the properties of matter can provide a clear evidence for changes
in the fundamental vacuum of QCD, with far-reaching consequences
\cite{Adams:2005dq}.

%%%%%%%%%%%%%%%%%%%     FIG. 1
%%%%%%%%%%%%%%%%%%%
%======================================================================
\section{The PNJL Model with Three Flavors}\label{formalism}

The Polyakov--Nambu--Jona-Lasinio (PNJL model) that we want to build
is an effective model for QCD, written in term of quark degrees of
freedom.  In the next paragraphs we will present the ingredients we
need to build our model namely: a massive Dirac Lagrangian together
with a four quark chirally invariant interaction (the original NJL
Lagrangian with a small mass term that breaks explicitly the chiral
symmetry, has observed in the mass spectrum); the so-called 't Hooft
interaction that reproduces the interaction of quarks with
instantons and finally a Polyakov loop potential that mimic the
effect of the pure gauge (Yang-Mills) sector of QCD on the quarks.

%======================================================================
\subsection{Nambu--Jona-Lasinio Model with Anomaly and Explicit Symmetry Breaking}

Phase transitions are usually characterized by large correlation
lengths, {\em i.e.}, much larger than the average distance between
the elementary degrees of freedom of the system. Effective field
theories then turn out to be a useful tool to describe a system near
a phase transition. In particular, in the usual Landau-Ginzburg
approach, the order parameter is viewed as a field variable and for
the latter an effective potential is built, respecting the
symmetries of the original Lagrangian. The existence of a phase
transition between two sectors where the chiral symmetry is
spontaneously broken or restored (transition associated to the quark
condensate that acts as an order parameter) and the Ginzburg--Landau
theory suggests to use the symmetry motivated NJL Lagrangian
\cite{Vogl:1991qt,Klevansky:1992qe,Hatsuda:1994pi,Buballa:2003qv}
for the description of the coupling between quarks and the chiral
condensate in the scalar-pseudoscalar sector. The associated
Lagrangian which complies with the underlying symmetries of QCD
described in the previous section reads:

\begin{eqnarray}
 \lagrange_{NJL} &=& \lagrange_0 + \lagrange_4 + \lagrange_6 \label{def:L_NJL} \\
 \lagrange_0 &=& \bar q(i \gamma^\mu \partial_\mu-\hat m)q \\
 \lagrange_4 &=& \frac{1}{2}\,g_S\,\,\sum_{a=0}^8\, [\,{(\,\bar q\,\lambda^a\, q\,)}
^2\,\,+\,\,{(\,\bar q \,i\,\gamma_5\,\lambda^a\, q\,)}^2\,] \\
 \lagrange_6 &=& g_D\,\{\mbox{det}\,[\bar q\,(1+\gamma_5)\,q] +\mbox{det}
 \,[\bar q\,(1-\gamma_5)\,q]\} %\nonumber\\
\end{eqnarray}

In the above $q = (u,d,s)$ is the quark field with three flavors
($N_f=3$) and three colors ($N_c=3$), $\hat{m} = \mbox{diag}(m_u,m_d,m_s)$ 
is the current quark mass matrix, and
$\lambda^a$ are the flavor SU$_f$(3) Gell-Mann matrices 
($a =0,1,\ldots , 8$), with ${ \lambda^0=\sqrt{\frac{2}{3}} \, {\bf I}}$.

As it is well known, this Lagrangian is invariant under a
global---and nonlocal---color symmetry $SU(N_c=3)$ and lacks the
confinement feature of QCD. \

$\lagrange_4$ has $U_L(3) \times U_R(3)$ chiral symmetry and
$\lagrange_6 $ is a six quarks interaction, the 't Hooft determinant
to reproduce $U_A(1)$ anomaly.  The $U_L(3) \times U_R(3)$ symmetry of
$\lagrange_0 + \lagrange_4$ is broken to $SU_L(3) \times SU_R(3)$ by
$\lagrange_6$.

The current quark mass matrix $\hat m$ is in general non degenerate
and explicitly breaks chiral symmetry $SU_L(3) \times SU_R(3)$ to
$SU_f(3)$ or its subgroup. In the following we will take $m_u = m_d$
hence keeping exact the isospin symmetry.

Let us notice that NJL has no build-in confinement and is
non-renormalizable thus it requires the introduction of a cutoff
parameter $\Lambda$.

The parameters,  obtained by following the methodology of
\cite{Rehberg:1995kh}, are: $m_u = m_d = 5.5$ MeV, $m_s =
140.7$ MeV, $g_S \Lambda^2 = 3.67$, $g_D \Lambda^5 = -12.36$ and
$\Lambda = 602.3$ MeV, which are fixed to reproduce the  values of
the coupling constant of the pion, $f_\pi\,=\,92.4$ MeV, and the
meson masses of the pion, the kaon, the $\eta$ and $\eta^\prime$,
respectively,  $M_\pi\,=\,135$ MeV, $M_K\,=\,497.7$ MeV,
$M_\eta\,=\,514.8$ MeV and $M_{\eta^\prime}\,=\,960.8$ MeV.

%======================================================================
\subsection{Coupling between Quarks and the Gauge Sector: The PNJL Model}

In the presence of dynamical quarks the $\Z_3$ symmetry is
explicitly broken: the expectation value of the Polyakov loop still
serves as an indicator for the crossover between the phase where
color confinement occurs ($\Phi \longrightarrow 0$) and the one
where color is deconfined ($\Phi \longrightarrow 1$).

The PNJL model attempts to describe in a simple way the two
characteristic phenomena of QCD, namely deconfinement and chiral
symmetry breaking.

In order to describe the coupling of quarks to the chiral condensate,
we start from a NJL description of quarks; they are coupled in a
minimal way to the Polyakov loop, via a covariant derivative.
Hence our calculations are performed in the framework of an extended
SU$_f$(3) PNJL Lagrangian, which includes the 't Hooft instanton
induced interaction term that breaks the U$_A$(1) symmetry and the
quarks are coupled to the (spatially constant) temporal background
gauge field $\Phi$
\cite{Fu:2007xc,Ciminale:2007sr,Fukushima:2008wg}. The Lagrangian
reads:
%
%%%%%%
\begin{eqnarray}\label{eq:lag}
{\mathcal L_{PNJL}\,}&=& \bar q(i \gamma^\mu D_\mu-\hat m)q +
\frac{1}{2}\,g_S\,\,\sum_{a=0}^8\, [\,{(\,\bar q\,\lambda^a\, q\,)}
^2\,\,+\,\,{(\,\bar q \,i\,\gamma_5\,\lambda^a\, q\,)}^2\,] \nonumber\\
&+& g_D\,\{\mbox{det}\,[\bar q\,(1+\gamma_5)\,q] +\mbox{det}
\,[\bar q\,(1-\gamma_5)\,q]\} %\nonumber\\
- \mathcal{U}\left(\Phi[A],\bar\Phi[A];T\right)
\end{eqnarray}
%%%%%%
%
The covariant derivative is defined as $D^{\mu}=\partial^\mu-i
A^\mu$, with $A^\mu=\delta^{\mu}_{0}A_0$ (Polyakov gauge); in
Euclidean notation $A_0 = -iA_4$.  The strong coupling constant $g$
is absorbed in the definition of $A^\mu(x) = g {\cal
A}^\mu_a(x)\frac{\lambda_a}{2}$, where ${\cal A}^\mu_a$ is the
(SU$_c$(3)) gauge field and $\lambda_a$ are the (color) Gell-Mann
matrices.

At $T=0$, it can be shown that the minimization of the grand
potential leads to $\Phi=\bar\Phi=0$. So, the quark sector decouples
from the gauge one, and the model is fixed as referred in the
previous subsection.

Some remarks are in order concerning the applicability of the PNJL
model. It should be noticed that in this model, beyond the chiral
point-like coupling between quarks,  the gluon dynamics is reduced
to a simple static background field representing the Polyakov loop
(see details in \cite{Ratti:2006gh,Hansen:2006ee}). This scenario is
expected to work only within a limited range of temperatures, since
at  large temperatures transverse gluons are expected to start  to
be thermodynamically active degrees of freedom and   they are not
taken into account in the PNJL model. We can assume that the range
of applicability of the model is roughly limited to $T\leq
(2-3)T_c$, since, as concluded in \cite{Meisinger:2003id},
transverse gluons start to contribute significantly for
$T>2.5\,T_c$, where $T_c$ is the deconfinement temperature.

%======================================================================
\subsection{Grand Potential in the Mean Field Approximation}

Our model of strongly interacting matter can simulate either a
region in the interior of neutron stars or a dense fireball created
in a heavy-ion collision. In the present work, we focus our
attention on the last type of systems.  Bearing in mind that in a
relativistic heavy-ion collision of duration of about $10^{-22}\,
s$, thermal equilibration is possible only for processes mediated by
the strong interaction rather than the full electroweak equilibrium,
we impose the condition $\mu_e=0$ and use the chemical equilibrium
condition $\mu_u=\mu_d=\mu_s=\mu$.
This choice allows for isospin symmetry, $m_u=m_d$ and approximates
the physical conditions at RHIC.

We will work in the (Hartree) mean field approximation. In this
context, the quarks can be seen as free particles whose bare current
masses $m_i$ are replaced by the constituent (or dressed) masses $M_i$.

The quark propagator in the constant background field $A_4$ is then:
\be
S_i(p) = -(\pslash - M_i+\gamma_0(\mu-iA_4))^{-1}
\ee
In the above, $p_0 = i \omega_n$ and $\omega_n =(2n+1)\pi
T$ is the Matsubara frequency for a fermion.

Within  the mean field approximation, it is straightforward
(see Ref.~\cite{Klevansky:1992qe}) to obtain  effective  quark masses
from the Lagrangian (\ref{eq:lag}); these masses are given by the
so-called gap equations:
%%%%%
\begin{equation}\label{eq:gap}
M_{i}=m_{i}-2g_{_{S}}\left\langle\bar{q_{i}}q_{i}\right\rangle
-2g_{_{D}}\left\langle\bar{q_{j}}q_{j}\right\rangle\left\langle\bar
{q_{k}}q_{k}\right\rangle\,
\end{equation}
%%%%%
where the quark condensates
$\left\langle\bar{q_{i}}q_{i}\right\rangle$, with $i,j,k=u,d,s$ (to
be fixed in cyclic order), have to be determined in a
self-consistent way as explained in the Appendix.

The PNJL grand canonical  potential density in the SU$_f$(3)
sector can be written as
%%%%%%
\be 
\Omega &=& \Omega(\Phi, \bar\Phi, M_i ; T, \mu) = {\cal
U}\left(\Phi,\bar{\Phi},T\right)
+ g_{_{S}}\sum_{\left\{i=u,d,s\right\}}\left\langle\bar{q_{i}}q_{i}\right\rangle^2 
+ 4 g_{_{D}}\left\langle\bar{q_{u}}q_{u}\right\rangle
\left\langle\bar{q_{d}}q_{d}\right\rangle\left\langle\bar{q_{s}}q_{s}\right\rangle \nonumber \\
&& - 2 N_c\sum_{\left\{i=u,d,s\right\}}
\int_\Lambda\frac{\mathrm{d}^3p}{\left(2\pi\right)^3}\,{E_i} 
- 2 T\sum_{\left\{i=u,d,s\right\}}\int\frac{\mathrm{d}^3{p}}{\left(2\pi\right)^3}
\left(z^+_\Phi(E_i) + z^-_\Phi(E_i) \right) \label{omega} 
\ee
%%%%%%
%
where  $E_i$ is the quasi-particle energy for the quark $i$:
$E_{i}=\sqrt{\mathbf{p}^{2}+M_{i}^{2}}$, and  $z^+_\Phi$ and $z^-_\Phi$ are the partition
function densities.
%%%%%

The explicit expression of $z^+_\Phi$ and $z^-_\Phi$ are given by:
\be z^+_\Phi(E_i) &\equiv& \mathrm{Tr}_c\ln\left[1+ L^\dagger \expp\right]= \ln\left\{ 1
+ 3\left( \bar\Phi + \Phi \expp \right) \expp
 + \expppp \right\} \label{eq:termo1}
\\
z^-_\Phi(E_i) &\equiv& \mathrm{Tr}_c\ln\left[1+ L \expm \right]= \ln\left\{ 1 + 3\left(
\Phi + \bar\Phi \expm \right) \expm
 + \expmmm \right\} \label{eq:termo2}
\ee where $E_i^{(\pm)}\,=\,E_i\,\mp \mu$, the upper sign applying
for fermions and the lower sign for anti-fermions. The technical
details are postponed to the Appendix but as a conclusion on this
topic, a word is in order to describe the role of the Polyakov loop
in the present model. Almost all physical consequences of the
coupling of quarks to the background gauge field stem from the fact
that in the expression of $z_\Phi$,  $\Phi$ or $\bar\Phi$ appear
only as a factor of the one- or two-quarks (or antiquarks) Boltzmann
factor, for example $\expm$ and $\expmm$. Hence when $\Phi,\bar\Phi
\rightarrow 0$ (signaling what we designate as the ``confined
phase'') only $\expmmm$ remains in the expression of the grand
potential, leading to a thermal bath with a small quark density. At
the contrary $\Phi,\bar\Phi \rightarrow 1$ (in the ``deconfined
phase'') gives a thermal bath with all 1-, 2- and 3-particle
contributions and a significant quark density. Other consequences of
this coupling of the loop to the 1- and 2-particles
Boltzmann factor will be discussed in Section \ref{CEP2}.

%======================================================================
\subsection{Equations of State and Response Functions}

The equations of state can be derived from the thermodynamical
potential $\Omega(T,\mu)$. This allows for the comparison of  some
of the results with  observables that have become accessible in
lattice QCD at non-zero chemical potential.

As usual, the pressure $p$ is defined such as its value is zero in the
vacuum state \cite{Buballa:2003qv} and, since the system is uniform, we have
%%%%%
\begin{equation} \label{press}
    p (T,\mu) = - \frac{1}{V}[\Omega(T,\mu)-\Omega(0,0)]
\end{equation}
%%%%%
where $V$ is the  volume of the system.

The relevant observables are the baryonic density
%%%%%
\begin{equation}\label{rho_B}
\rho_B(T,\mu) = -\frac{1}{V} \left (\frac{\partial\Omega}{\partial\mu}\right )_T
\end{equation}
%%%%%
and the (scaled) ``pressure difference'' given by
%%%%%
\begin{equation}
\frac{\Delta p(T,\mu)}{T^4}=\frac{p(T,\mu)-p(T,0)}{T^4}
\end{equation}
%%%%%

Due to the relevance for the study of the thermodynamics of matter
created in relativistic heavy-ion collisions, it is interesting to
perform an analysis of the isentropic trajectories.
The equation of state for the entropy density, $s$, is given by
%%%%%
\begin{equation}\label{entropy}
s(T,\mu)\,= \left (\frac{\partial p}{\partial T}\right)_{\mu}\,
\end{equation}
%%%%%
and  the  energy density, $\epsilon$, comes from the following fundamental relation of
thermodynamics

%%%%%
\begin{equation}\label{energydens}
    \epsilon (T, \mu)\,=\,T\,s(T,\mu)\,+\,\mu\,\rho_B(T,\mu)\,-\,p(T,\mu)\,
\end{equation}
%%%%%

The energy density and the pressure are defined such that their values are zero in the vacuum state
\cite{Buballa:2003qv}.

The baryon number susceptibility, $\chi_B$, and the specific heat,
$C$, are  the response of the baryon number density,
$\rho_B(T,\mu)$, and the entropy density, $s (T,\mu)$, to an
infinitesimal variation of the quark chemical potential $\mu$ and
temperature, given respectively by:
%%%%%
\begin{equation} \label{chi}
    \chi_B = \left(\frac{\partial \rho_B}{\partial\mu}\right)_{T} \hskip1cm  {\rm and}
    \hskip1cm  C = \frac{T}{V}\left ( \frac{\partial s}{\partial T}\right)_{\mu}
\end{equation}
%%%%%

These second order derivatives of the pressure are relevant
quantities to discuss phase transitions, mainly the second order
ones.

%======================================================================
\subsection{Model Parameters and Regularization Procedure}

As already seen, the pure NJL sector involves five parameters: the
coupling constants $g_S$ and $g_D$, the current quark masses
$m_u=m_d$, $m_s$ and the cutoff $\Lambda$ defined in Section
\ref{formalism}. These parameters are determined in the vacuum by
fitting the experimental values of several physical quantities.
We notice that the coupling constant $g_S$ and $g_D$ and the
parameter $\Lambda$ are correlated. For instance, if we increase
$g_S$ in order to provide a more significant attraction between
quarks, we must also increase the cutoff $\Lambda$ in order to
insure a good agreement with experimental results. In addition, the
value of the cutoff itself does have some impact as far as the
medium effects in the limit $T=0$ are concerned. Also the strength
of the coupling $g_D$ has a relevant role in the location of the
CEP, as it will be discussed.

In fact, the choice of the parametrization may  give rise to
different physical scenarios at $T=0$ and $\mu_B=\mu\neq 0$
\cite{Buballa:2003qv}, even if they give reasonable fits to hadronic
vacuum observables  and predict a first order phase transition.
The set of parameters we used insures the stability conditions and,
consequently, the compatibility with  thermodynamic expectations.

On the other hand, the  regularization procedure, as soon as the
temperature effects are considered, has relevant consequences on the
behavior of physical observables, namely on the chiral condensates
and the meson masses \cite{Costa:2007fy}. Advantages and drawbacks
of these regularization procedures have been discussed within the
NJL \cite{Costa:2007fy} and PNJL \cite{Costa:2009ae,Ruivo:2010fc}
models.  We remind that one of the drawbacks of the regularization
that consists in putting a cutoff only on the divergent integrals 
is that, at high temperature, there is a too fast decrease
of the quark masses that  become lower than their current values.
This leads to a non physical behavior of the quark condensates that,
after vanishing at the temperature ($T_{eff}$) where constituent and
current quark masses coincide, change sign and acquire non-zero
values again. To avoid this unphysical effects  we use the
approximation of imposing by hand the condition that, above
$T_{eff}$, $M_i=m_i$ and  $\left\langle \bar{q_i}
q_i\right\rangle=0$.

%======================================================================
%======================================================================
\section{Equation of State at Finite Temperature}\label{P_T_finite_T}

%======================================================================
\subsection{Characteristic Temperatures}

At zero temperature and chemical potential, the chiral symmetry of
QCD is explicitly broken. It is expected that chiral symmetry will
be restored at high temperature; hence a phase transition occurs
separating the low and the high temperature regions. This phenomena
may be realized in high energy heavy ion collision experiments.

In this case ($\mu_B=0$ and $T\neq 0$), where a crossover transition
occurs in the PNJL model, a unique critical temperature cannot be
defined, but one of many characteristic temperatures for the
transition may be used. Of course, these temperatures should
coincide in the limit where the transition becomes of second order
(in the chiral limit, for example, when one is concerned with the
restoration of chiral symmetry).

We start our analysis by identifying the characteristic temperatures
which separate the different thermodynamic phases in the PNJL model
\cite{Hansen:2006ee}, using the regularization that allows high
momentum quark states at finite temperature (the cutoff is used
only in the integrals that are divergent and $\Lambda \rightarrow
\infty$ in the ones that are already convergent because of the
thermal distributions).

Let us analyze the behavior of  the quark masses and the field
$\Phi$. The characteristic temperature related to the deconfinement
phase transition is $T_c^\Phi$ and the chiral phase transition
characteristic temperature, $\Tcc$, signals partial restoration of
chiral symmetry. These temperatures are chosen to be, respectively,
the inflexion points  of the ``quasi'' order parameter $\Phi$ and of
the chiral condensate $\langle\bar{q_u}q_u\rangle$; as in
\cite{Ratti:2006gh}, we define $T_c$ as the average of the two
transition temperatures, $T_c^\Phi$ and $\Tcc$. As shown in
\cite{Ruivo:2010fc}, the present regularization lowers the
characteristic temperatures and decreases the gap $T_c^\Phi-\Tcc$,
leading therefore to better agreement with lattice results.

Let us remark that this section is also devoted to the study of
thermodynamic quantities (pressure, energy per particle and entropy)
for which a rescaling of the temperature $T_0$ is needed, in order
to get a better agreement with lattice results for these quantities.
Therefore, following the argumentation presented in
\cite{Ratti:2006gh}, here we will use the reduced temperature $T_c$
by rescaling the parameter $T_0$ from 270 to 210 MeV (let us stress
that this rescaling is only done for the remainder of this section).
Results for the characteristic temperatures with $T_0=210$ MeV and
$T_0=270$ MeV are shown in Table \ref{table:TcT0}.

%%%%%%%%%%%%%%%%%%%%%%%%%%%%%%%%%%%%
\begin{table}[t]
\begin{center}
    \begin{tabular}{|c|c|c|c|}
        \hhline{|----|}
        $T_0$ [MeV] & $T^\chi_c$ [MeV] &  $T^\Phi_c$ [MeV] & $T_c$ [MeV] \\
        \hline\hline
        270 & 222 & 210 & 216 \\
        \hhline{|----|}
        210 & 203 & 171 & 187\\
        \hhline{|----|}
    \end{tabular}
\caption{\label{table:TcT0} Characteristic temperatures with two different values for $T_0$.}
\end{center}
\end{table}
%%%%%%%%%%%%%%%%%%%%%%%%%%%%%%%%%%%%

It should be noticed that, when  $T_0=210$ MeV is used, we loose the
almost perfect coincidence of the chiral and deconfinement
transitions (they are shifted relative to each other by about 32
MeV) and we have  $T_c = 187$ MeV within the range expected from
lattice calculations \cite{Cheng:2009zi}. However, the behavior of
the relevant physical quantities is qualitatively the same whether
$T_0=270$ MeV or $T_0=210$ MeV.

\begin{figure}[t]
\begin{center}
\vspace{-0.5cm}
  \begin{tabular}{cc}
    \epsfig{file=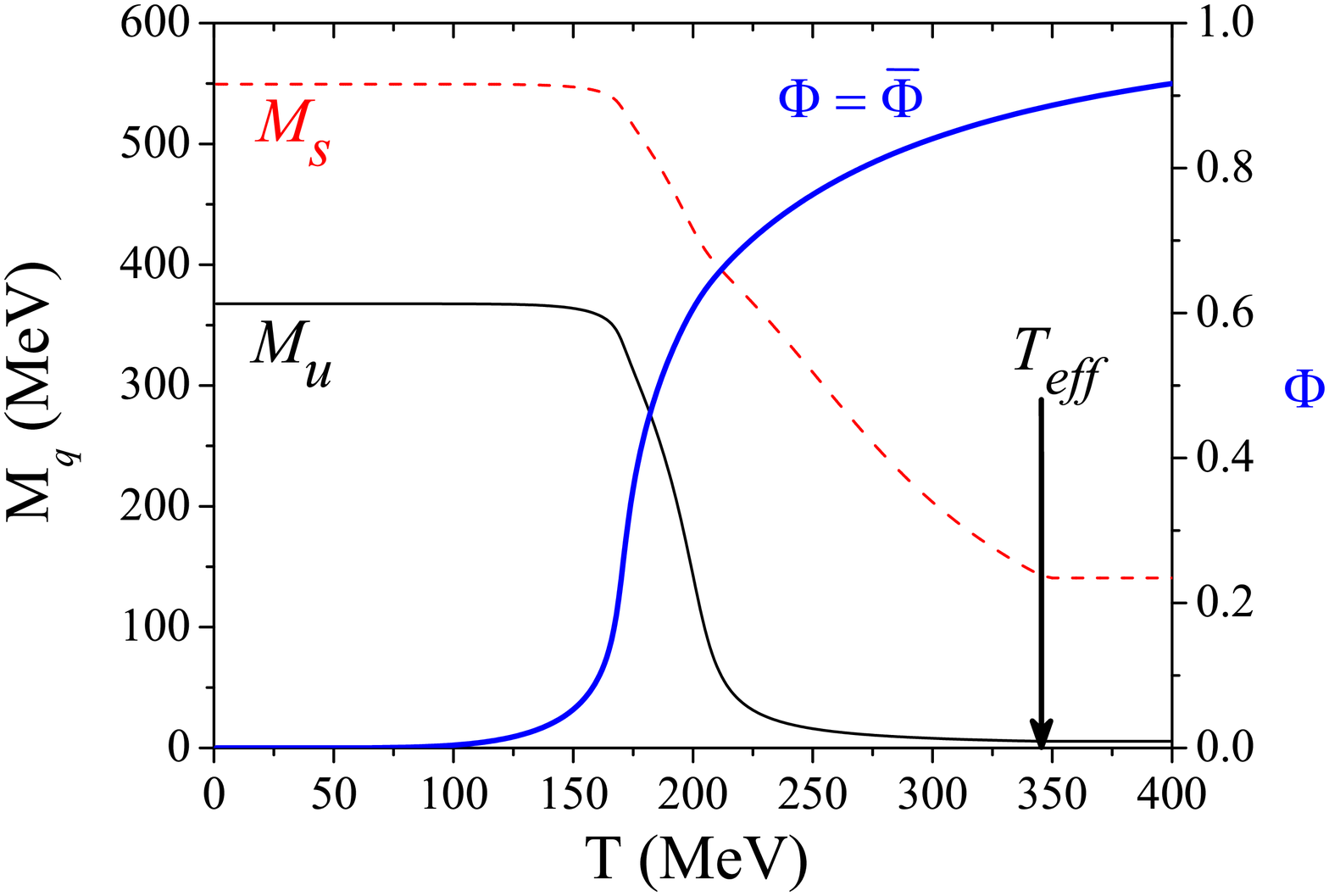,width=8cm,height=7cm} &
    \epsfig{file=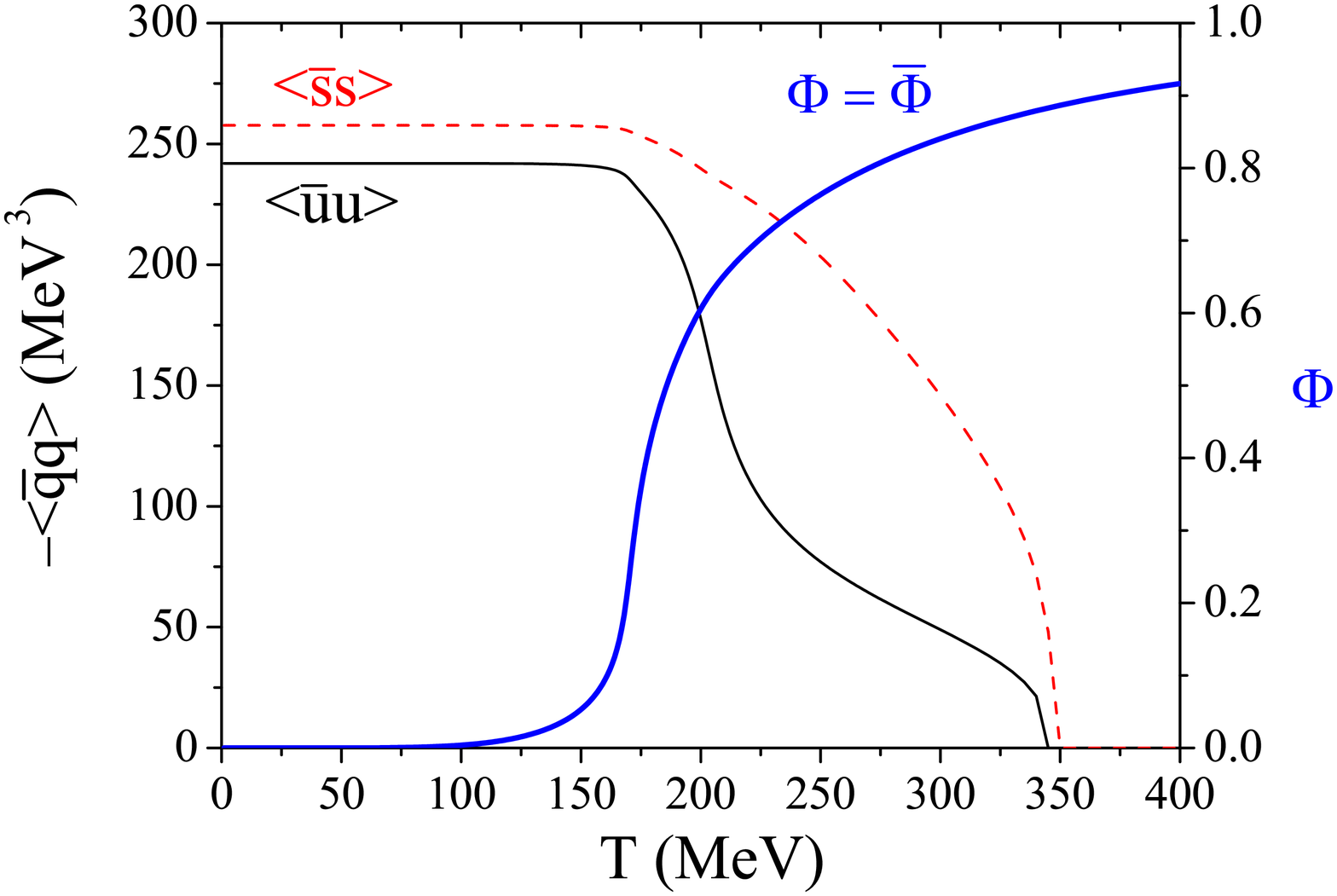,width=8cm,height=7cm} \\
    \epsfig{file=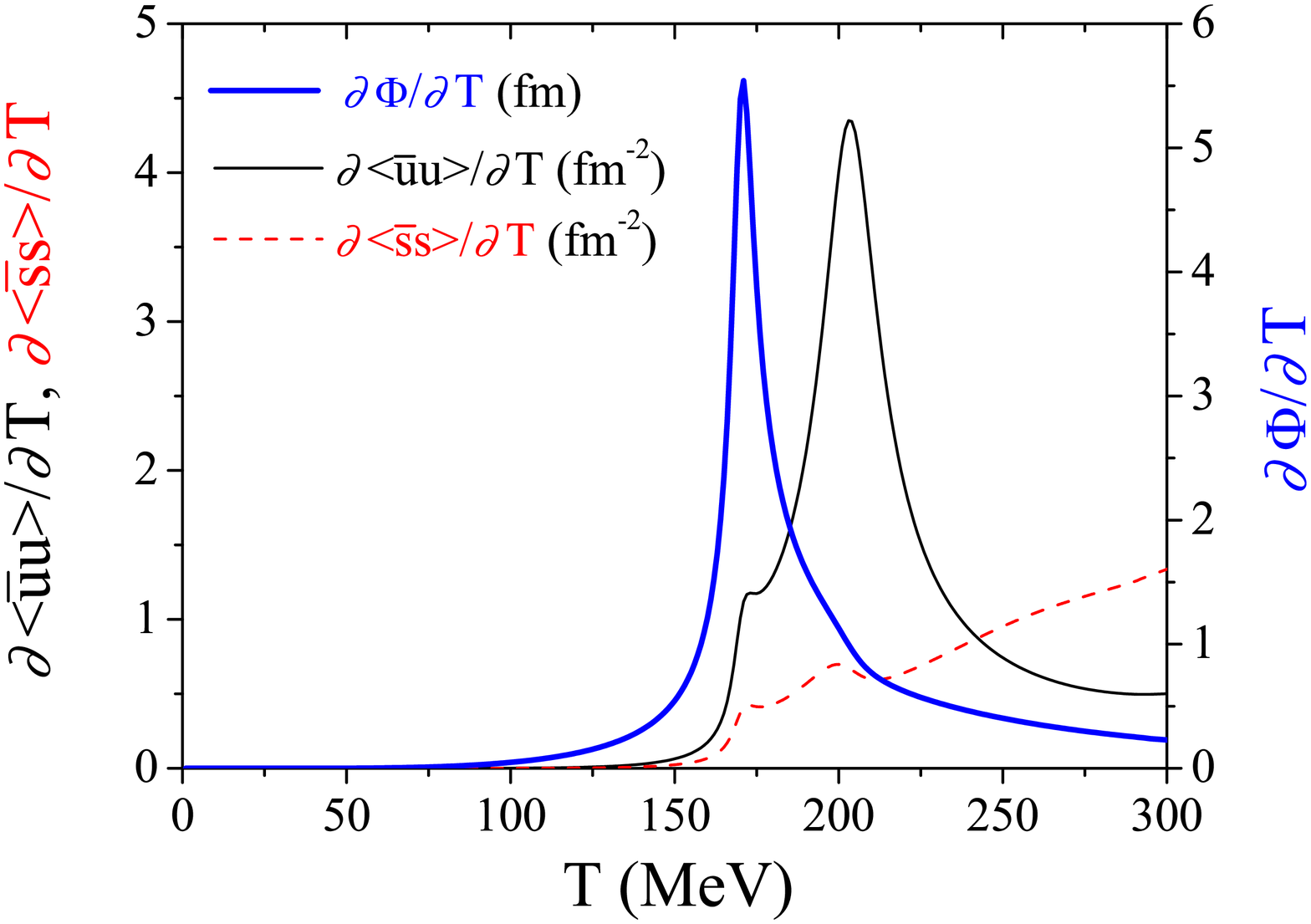,width=8cm,height=7cm} &
    \epsfig{file=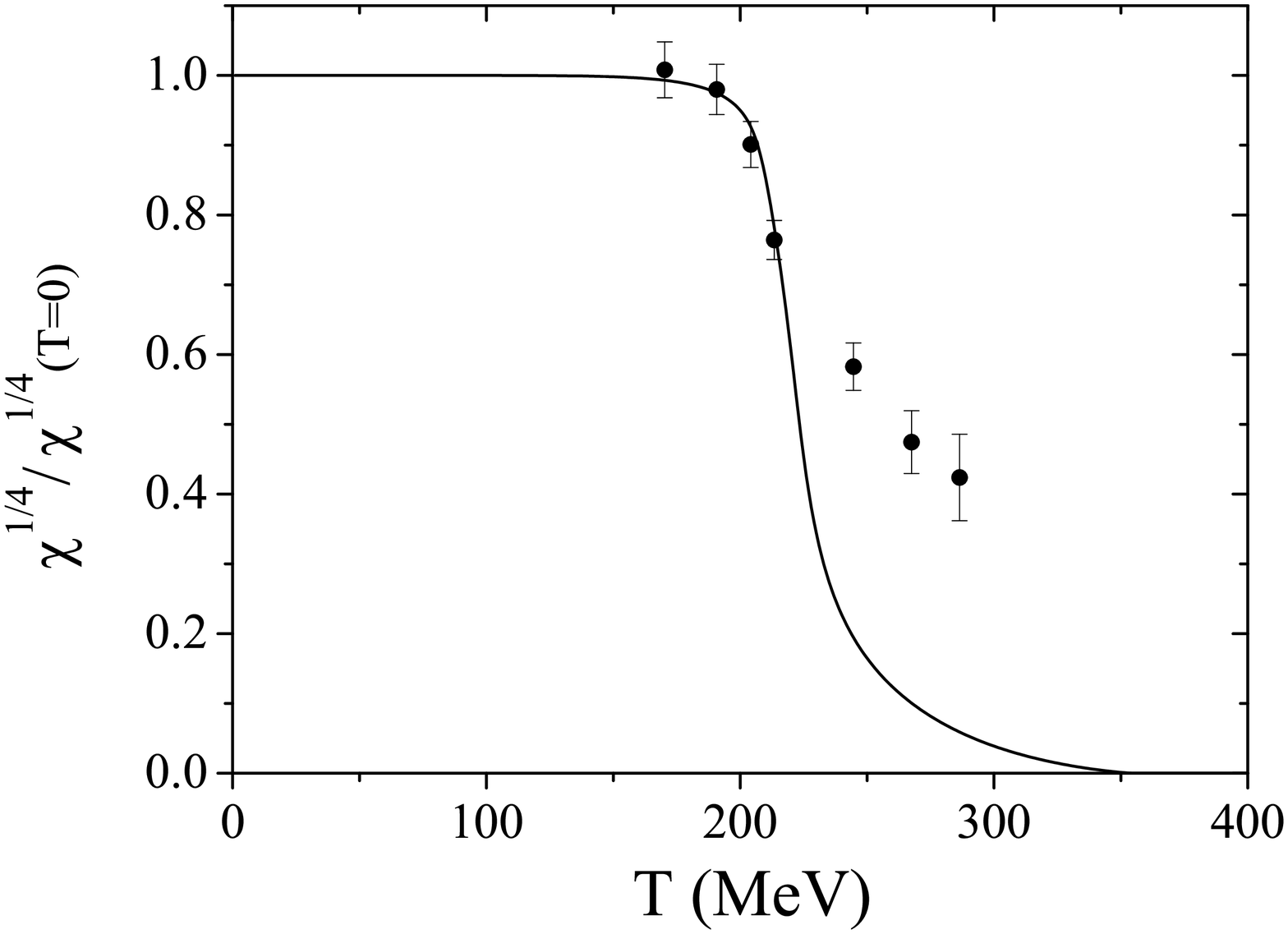,width=8cm,height=7cm} \\
   \end{tabular}
\vspace{-0.5cm}
\end{center}
\caption{ Upper part: quark masses (left panel) and quark condensates
(right panel) in PNJL model as functions of the temperature; 
the Polyakov loop field $\Phi$ is also shown.
At $T_{eff} = 345$ MeV, $M_i = m_i$. Lower part: derivatives
of the quark condensates and of the Polyakov loop field $\Phi$ (left panel);
the topological susceptibility, $\chi$, (right panel) in the PNJL model,
compared to corresponding lattice results taken from \cite{Alles:1996nm}.}
\label{Fig:massas}
\end{figure}

An interesting point to be noticed in Figure \ref{Fig:massas},
lower panel, is that the inflexion points of the strange and
non-strange quark condensates are close and slightly separated from
the inflexion point of $\Phi$, although the condensates have small
peaks around the inflexion point of $\Phi$. 
%%%%%%%%%%%%%%%%%%  FIG. 2
%%%%%%%%%%%%%%%%%%%%
The latter effect is
related to the fastening of the phase transition induced by the
Polyakov loop as shown in \cite{Costa:2008dp} (it obviously does not
signal a phase transition since the variation of the order parameter
around this temperature is small). The Polyakov loop leads to a
faster decrease of the quark masses around $T_c$ and the present
regularization enhances this effect, even at temperatures higher
than $T_c$. At $T_{eff}=345$ MeV the regularization is responsible
for the full restoration of the chiral symmetry that was dynamically
broken: the quark masses go to their current values and the quark
condensates vanish (see Figure \ref{Fig:massas}, left and right
panels). As already shown in the framework of the pure NJL model
\cite{Costa:2007fy} and in the PNJL model \cite{Ruivo:2010fc}, the
effect of allowing high momentum quark states is stronger for the
strange quark mass. Indeed, with the conventional regularization,
the non-strange constituent quark mass at high temperature is
already very close to its current mass, the new regularization only
enhancing this behavior. Differently, the strange quark mass is
always far from its current value, unless we allow high momentum
quarks to be present and, in that case, its constituent mass
decreases very substantially and comes to its current value.

As shown in \cite{Ruivo:2010fc}, at $T_{eff}$ the behavior of some
given observables signals the effective restoration of chiral and
axial symmetry: the masses of the meson partners of both chiral and
axial symmetry are degenerated and the topological susceptibility
vanishes as we can see from Figure \ref{Fig:massas} (lower right
panel).

%======================================================================
\subsection{Thermodynamic Quantities}

In the limit of vanishing quark chemical potential, significant
information on the phase structure of QCD at high temperature is
obtained from lattice calculations. The transition to the phase
characteristic of this regime is related with chiral and
deconfinement transitions which are the main features of our model
calculation.

\begin{figure}[t]
\begin{center}
\vspace{-0.5cm}
  \begin{tabular}{cc}
    \epsfig{file=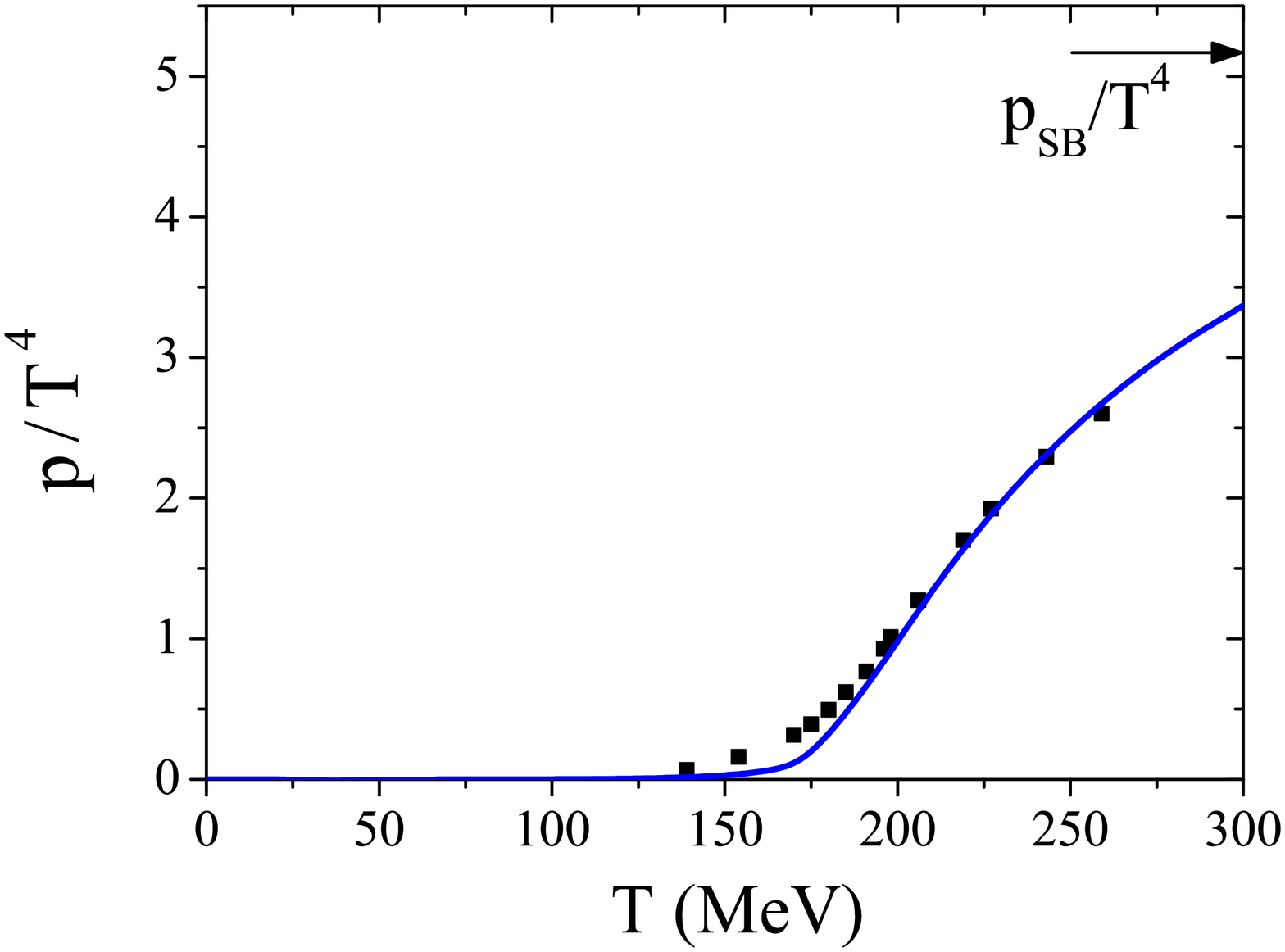,width=9cm,height=7.5cm} &
    \hspace*{-1cm}\epsfig{file=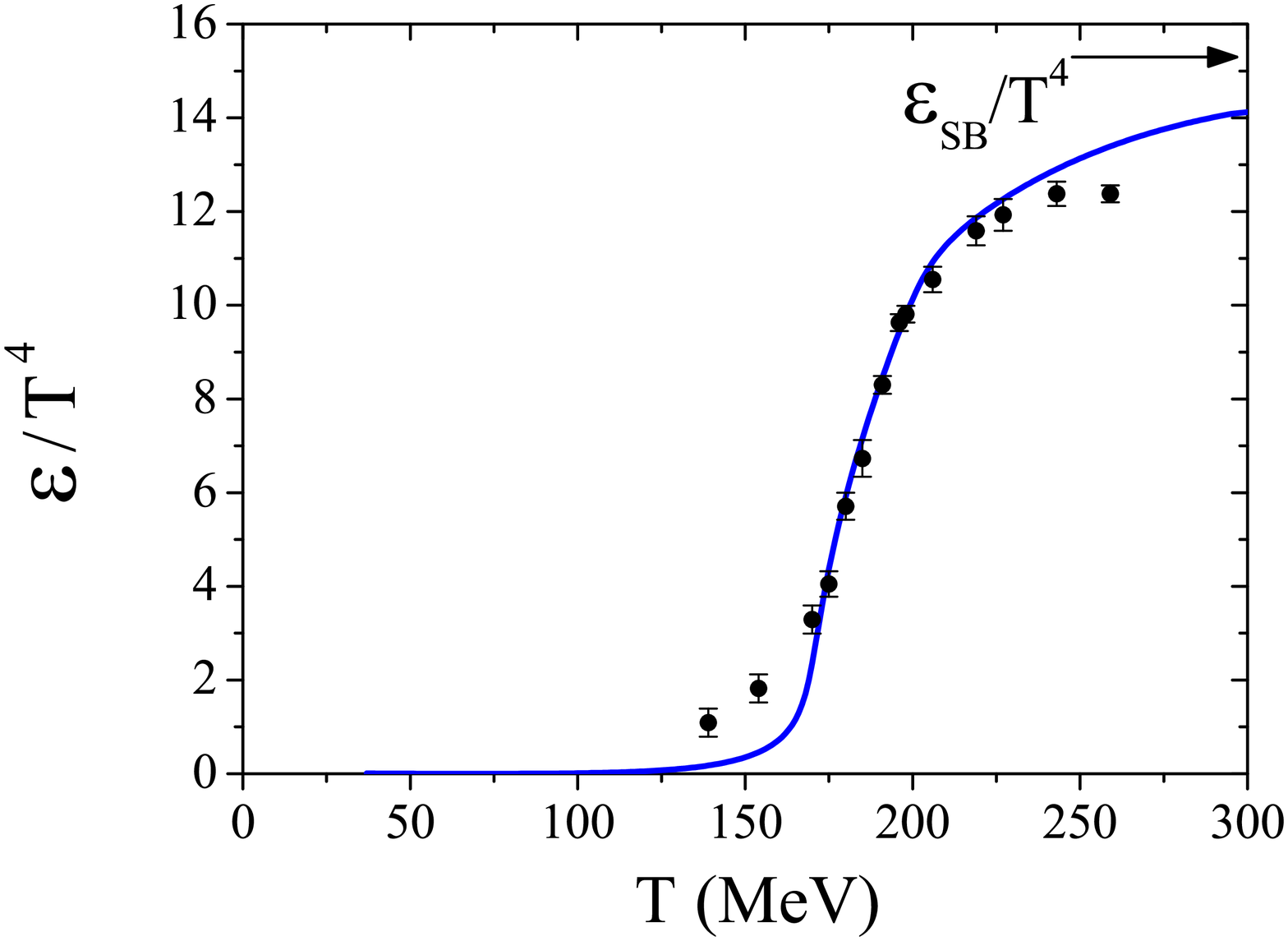,width=9cm,height=7.5cm} \\
   \end{tabular}
  \epsfig{file=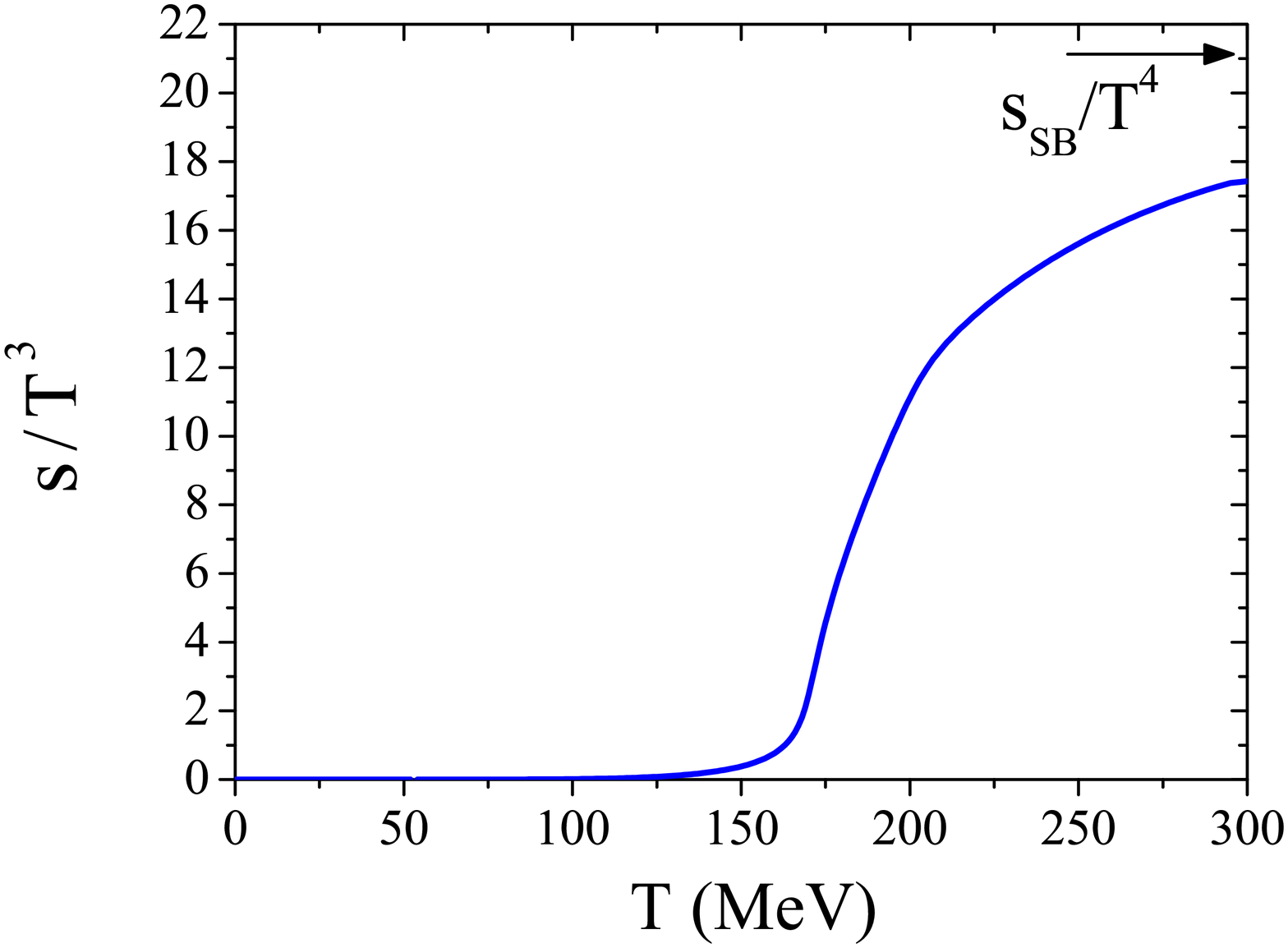,width=9cm,height=7.5cm}
\end{center}
\caption{Scaled pressure $(p)$,  energy  per particle $(\epsilon)$, and
entropy $(s)$ as a function of the temperature at zero chemical potential. The data
points are taken from \cite{Cheng:2009zi}. The pressure reaches 66\% of the strength of
the Stefan-Boltzmann value at $T=300$ MeV, a value which attains 85\% at $T=400$ MeV.}
\label{Fig:termod}
\end{figure}

In Figure \ref{Fig:termod}, we plot the scaled pressure, the energy
and the entropy as functions of the temperature compared with recent
lattice results (see Reference \cite{Cheng:2009zi}). Since the
transition to the high temperature phase is a rapid crossover rather
than a phase transition, the pressure, the entropy and the energy
densities are continuous functions of the temperature. We observe a
similar behavior in the three curves: a sharp increase in the
vicinity of the transition temperature and then a tendency to
saturate at the corresponding ideal gas limit. Asymptotically, the
QCD pressure for $N_f$ massless quarks and $(N_c^2 - 1)$ massless
gluons is given  ($\mu_B=0$)  by:
%%%%%%%
\begin{equation}\label{pSB}
\frac{p_{SB}}{T^4}\,=\,(N_c^2 - 1)\,\frac{\pi^2}{45}\,+\,N_c\,N_f\,\frac{7\,\pi^2}{180},
\end{equation}
%%%%%%%
where the first term denotes the gluonic contribution and the second
term the fermionic one.

%%%%%%%%%%%%%%%%%%  FIG. 3
%%%%%%%%%%%%%%%%%%%%

The results follow the expected tendency and go to the free gas
values (or Stefan--Boltzmann limit), a feature that was also found
with this type of regularization in the context of the SU(2) PNJL
model \cite{Megias:2004hj,Megias:2006bn,Ratti:2006gh}.
The inclusion of the Polyakov loop effective potential ${\cal
U}(\Phi,\bar\Phi;T)$ (it can be seen as an effective pressure term
mimicking the gluonic degrees of freedom of QCD) is required to get
the correct limit (indeed in the NJL model the ideal gas limit is
far to be reached due to the lack of gluonic degrees of freedom).

The inclusion of the Polyakov loop and the regularization procedure
are essential to obtain the required increase of extensive
thermodynamic quantities, insuring the convergence to the
Stefan-Boltzmann (SB) limit of QCD \cite{Philipsen:2007rj}. Some
comments are in order concerning the role of the regularization
procedure for $T>T_c$. In this temperature range, due to the
presence of high momentum quark states, the physical situation is
dominated by the significant decrease of the constituent quark
masses by the $q \bar q$ interactions.  This allows for an ideal gas
behavior of almost massless quarks with the correct number of
degrees of freedom.

Let us notice that, just below $T_c$, the pressure and the energy
fail to reproduce the lattice points: for example there is a small
underestimation of the pressure and energy in the model
calculations. It is known that the lack of mesonic correlations in
the PNJL model is responsible for, at least, a fraction of this
discrepancy. As matter of fact, in \cite{Blaschke:2007np} the
authors have evaluated the pion and sigma contributions to the
pressure by calculating the ring sum in the SU(2) nonlocal PNJL
model. They have shown that the pionic contribution dominates the
pressure in the low temperature region, shifting the respective
curve to higher values. It is expected that the introduction of
mesonic correlations in the SU(3) PNJL model has the same effect
for the pressure.

%======================================================================
%======================================================================
\section{Phase Diagram and the Location of the Critical End Point}\label{phase}

Although recent lattice QCD results by de Forcrand and Philipsen
question the existence of the CEP
\cite{deForcrand:2006pv,deForcrand:2007rq,deForcrand:2008vr}, this
critical point of QCD, proposed at the end of the eighties
\cite{Asakawa:1989bq,Barducci:1989wi,Barducci:1989eu,Barducci:1993bh},
is still a very important subject of discussion nowadays \cite{CEP}.

We remember that the TCP separates the second order transition at
high temperature and low chemical potential, from the first order
transition at high chemical potential and low temperature. If the
second order transition is replaced by a smooth crossover, a CEP
which separates the two lines is found. In order to determine and
elucidate the nature of the phase transition the relevant
thermodynamic quantities are studied, starting with zero temperature
and  finite chemical potential.

%======================================================================
\subsection{Phase Transition at Zero Temperature}

\begin{figure}[t]
\begin{center}
\vspace{-0.5cm}
  \begin{tabular}{cc}
    \epsfig{file=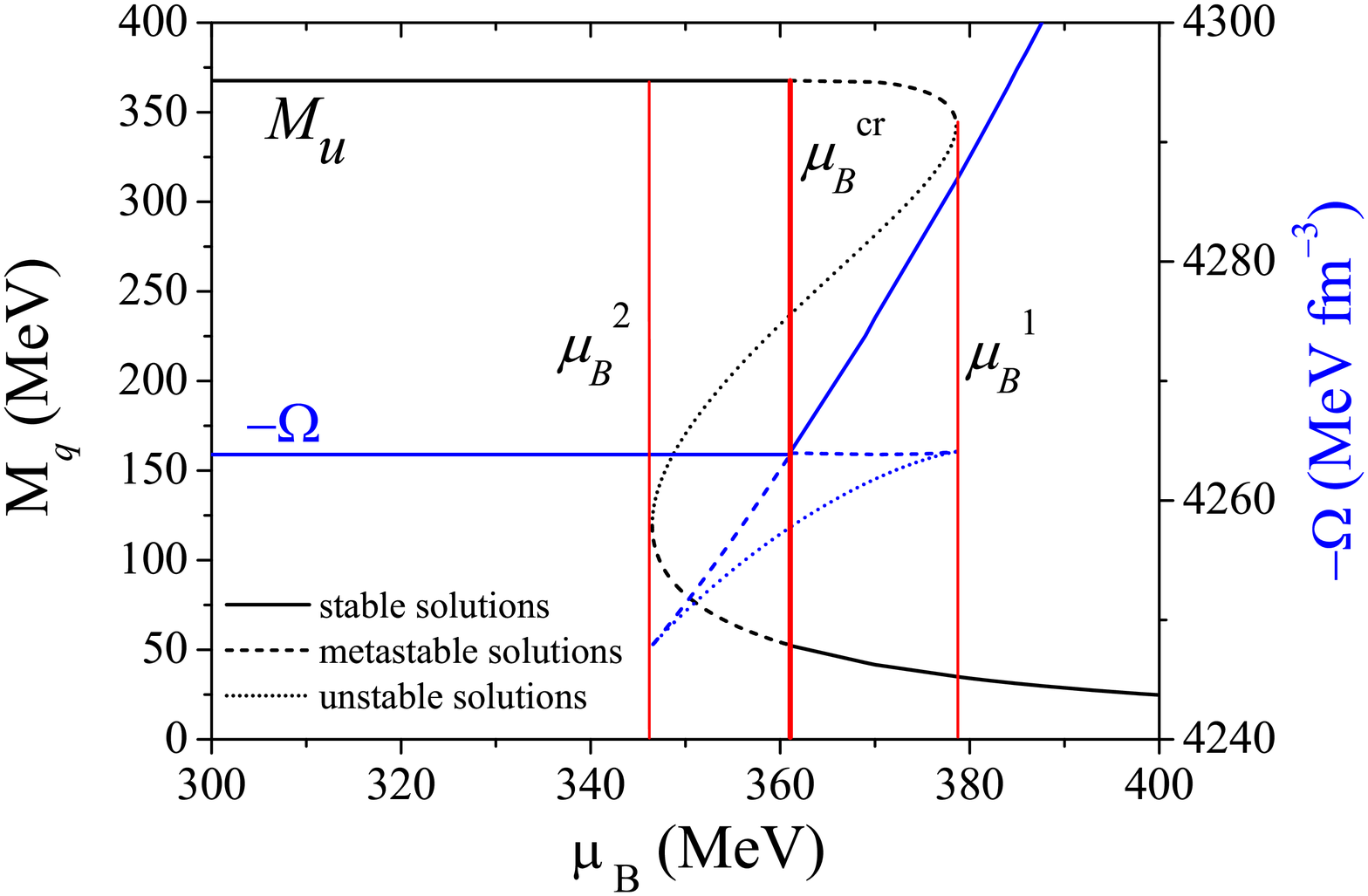,width=8.5cm,height=7.5cm} &
    \epsfig{file=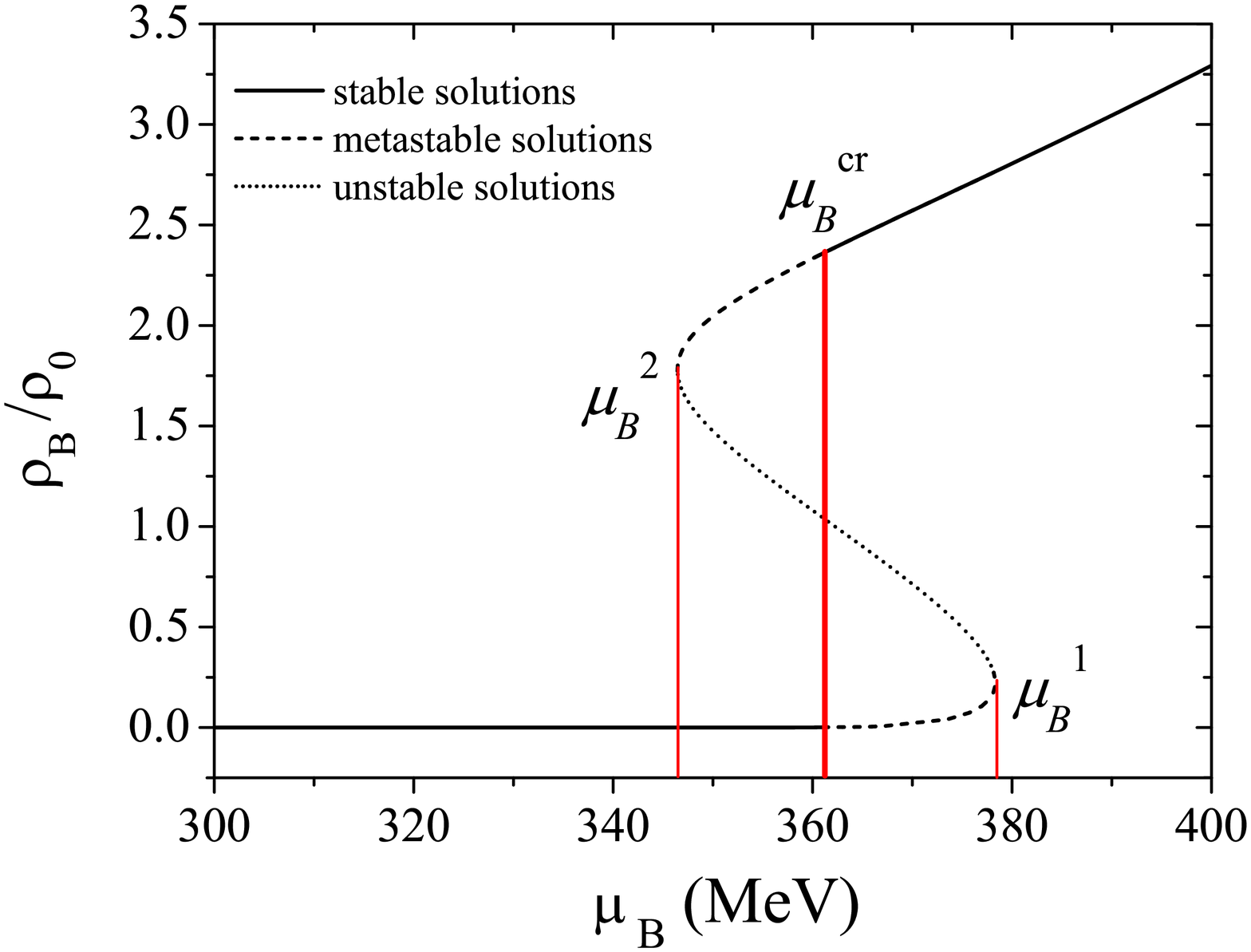,width=8.5cm,height=7.5cm} \\
   \end{tabular}
\end{center}
\vspace{-0.5cm}
\caption{
Left: grand potential for the stable, metastable and unstable
phases, used to determine $\mu_B^ {cr}$ and $u$ quark mass; right: baryonic density as
function of $\mu_B$.}
\label{Fig:massa_dens}
\end{figure}

Figure \ref{Fig:massa_dens} illustrates the properties of cold quark
matter as predicted by the model calculation. In the left panel  the
negative of the thermodynamic potential is plotted as a function of
the chemical potential,  showing the presence of branches with
stable, metastable and unstable solutions. This behavior is obtained
in the  domain $\mu_B^2<\mu_B<\mu_B^1$, where the gap equations
(\ref{eq:gap}) have three solutions as illustrated in the left panel
of Figure \ref{Fig:massa_dens} for the constituent non-strange quark
mass.
As a consequence, a first order phase transition is found at the
critical chemical potential $\mu_B^{cr}= 361.7$ MeV.
The stable solutions are realized by the minimum of the
thermodynamic potential. When stable and metastable solutions give
the same value for the thermodynamic potential, the phase transition
occurs as illustrated in Figure \ref{Fig:massa_dens} (left panel).
The phase of broken symmetry is realized for $\mu_B<\mu_B^{cr}$ and
the ``symmetric'' phase is realized for $\mu_B>\mu_B^{cr}$. At this
crossing point of the curve, the two phases are in thermal and
chemical  equilibrium (Gibbs criteria). The baryon density,
represented in Figure \ref{Fig:massa_dens} (right panel) as function
of $\mu_B$, is given by the slope of the curve $- \Omega$ as
indicated by Equation (\ref{rho_B}). This quantity is plotted in
Figure \ref{Fig:massa_dens} (right panel) which shows that the
condition of thermodynamic stability $\left({\partial (-
\Omega)}/{\partial \mu_B}>0\right)$ is violated by the portion of
the curve with negative curvature (unstable phase).

%%%%%%%%%%%%%%%%%%%%    FIG. 4
%%%%%%%%%%%%%%%%%%%%

\begin{figure}[t]
\begin{center}
  \vspace*{-0.4cm}
  \begin{tabular}{cc}
    \hspace*{-0.5cm}\epsfig{file=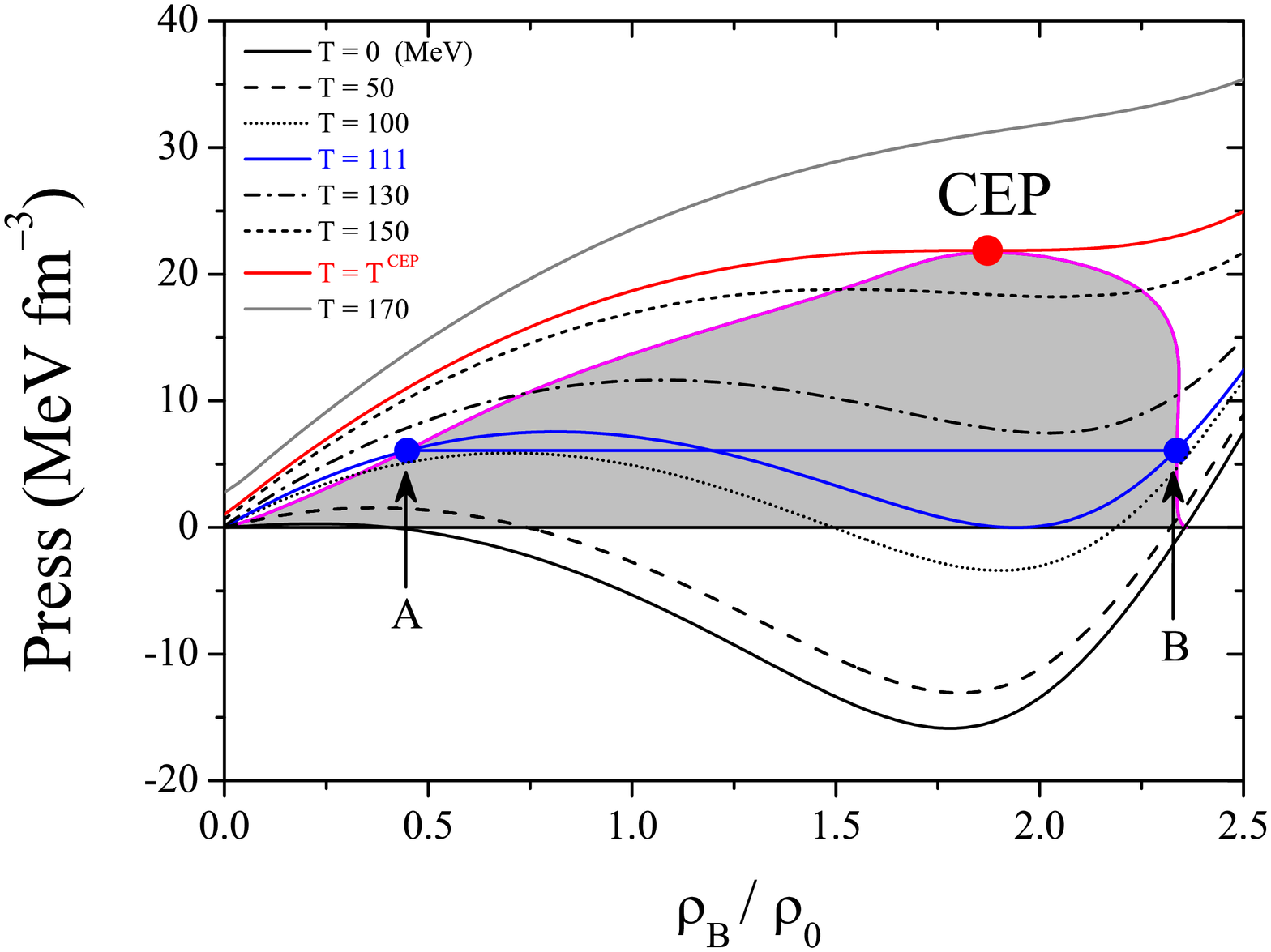,width=9cm,height=7.5cm} &
    \hspace*{-1cm}\epsfig{file=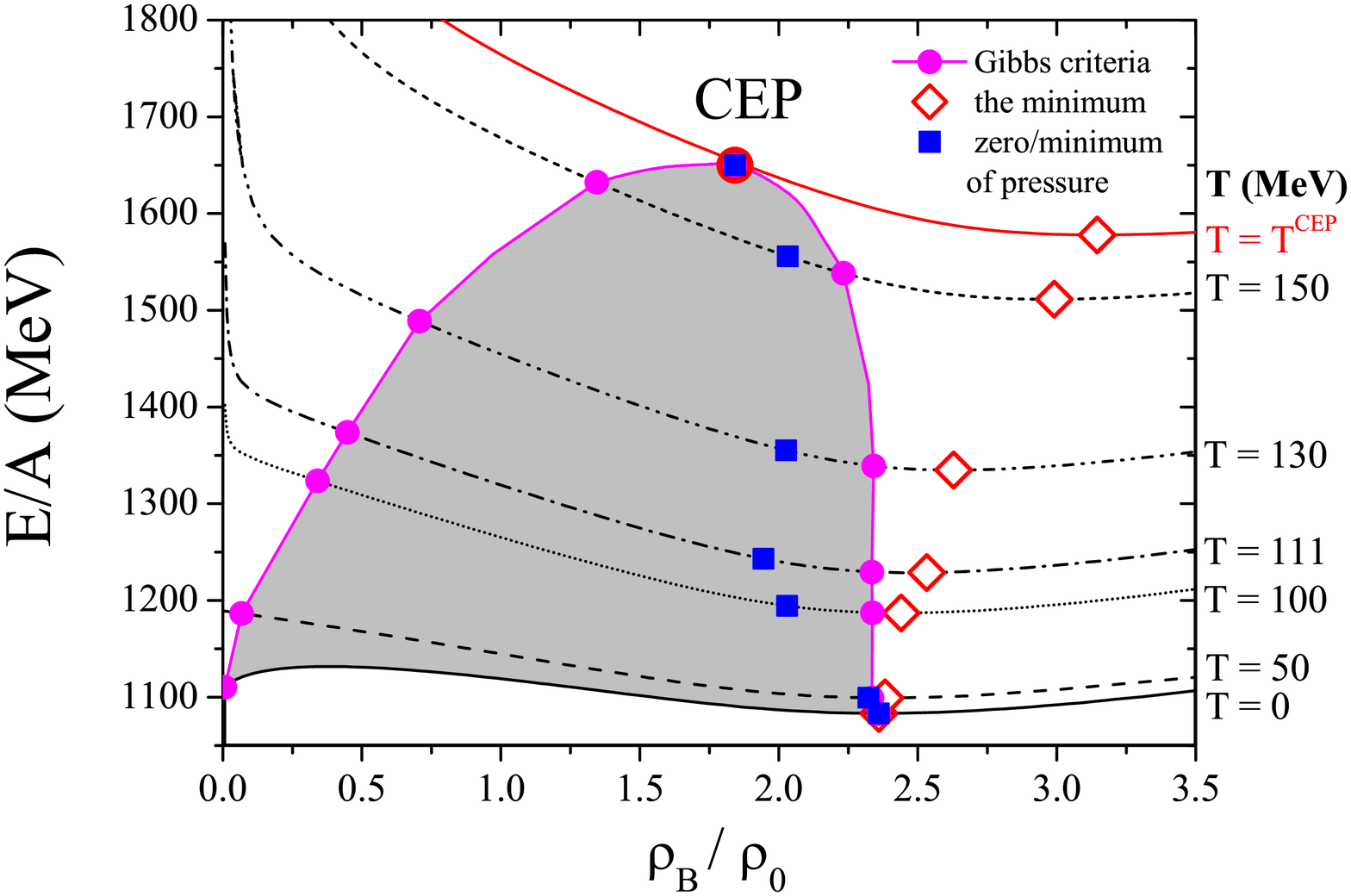,width=9cm,height=7.5cm}
   \end{tabular}
\end{center}
\vspace*{-0.5cm}
\caption{
Pressure (left) and energy per particle (right) as a function of
$\rho_B/\rho_0$ at different temperatures ($\rho_0 = 0.17$ fm$^{-3}$ is the normal
nuclear matter density). The points $A$ and $B$ (left panel) illustrate the Gibbs
criteria. Only in the $T=0$ line the zero-pressure point is located at the minimum of
the energy per particle.}
\label{Fig:press_energ}
\end{figure}

This information can be  complemented by the behavior of the
pressure/energy per particle as a function of the baryonic density.
To this purpose let us analyze the curve at $T=0$  in Figure
\ref{Fig:press_energ}. The pressure has three zeros that correspond
to the extrema of the energy per particle. The third zero of the
pressure, at $\rho_B=2.36 \rho_0$, corresponds to an absolute
minimum of the energy (see Figure \ref{Fig:press_energ} right
panel).
This is an important point of the  model calculation, and the set of
parameters is chosen in order to insure such a condition. In fact,
the occurrence of an absolute minimum of the energy allows for the
existence of finite droplets in mechanical equilibrium with the
vacuum at zero--pressure ($P=0$). For densities above a critical
value, $\rho_B^{cr}=2.36 \rho_0$, the system returns to a uniform
gas phase. The equilibrium configuration for densities
$0<\rho_B<\rho_B^{cr}$ is, therefore, a mixed phase.

%%%%%%%%%%%%%%%%%%  FIG. 5
%%%%%%%%%%%%%%%%%%%%

\begin{figure}[t]
\begin{center}
  \vspace*{-0.5cm}
  \begin{tabular}{cc}
    \epsfig{file=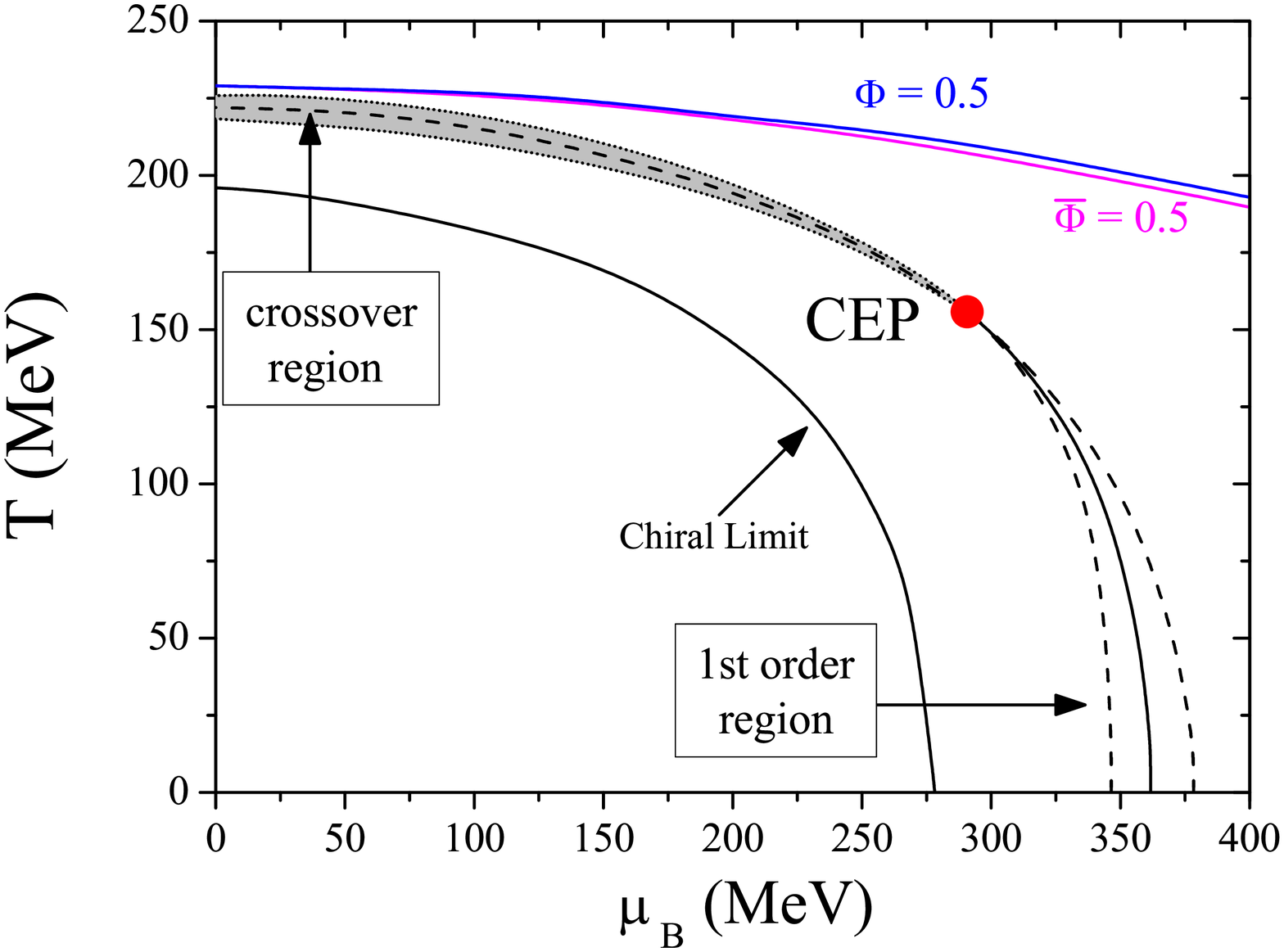,width=9cm,height=7.5cm} &
    \hspace*{-1cm}\epsfig{file=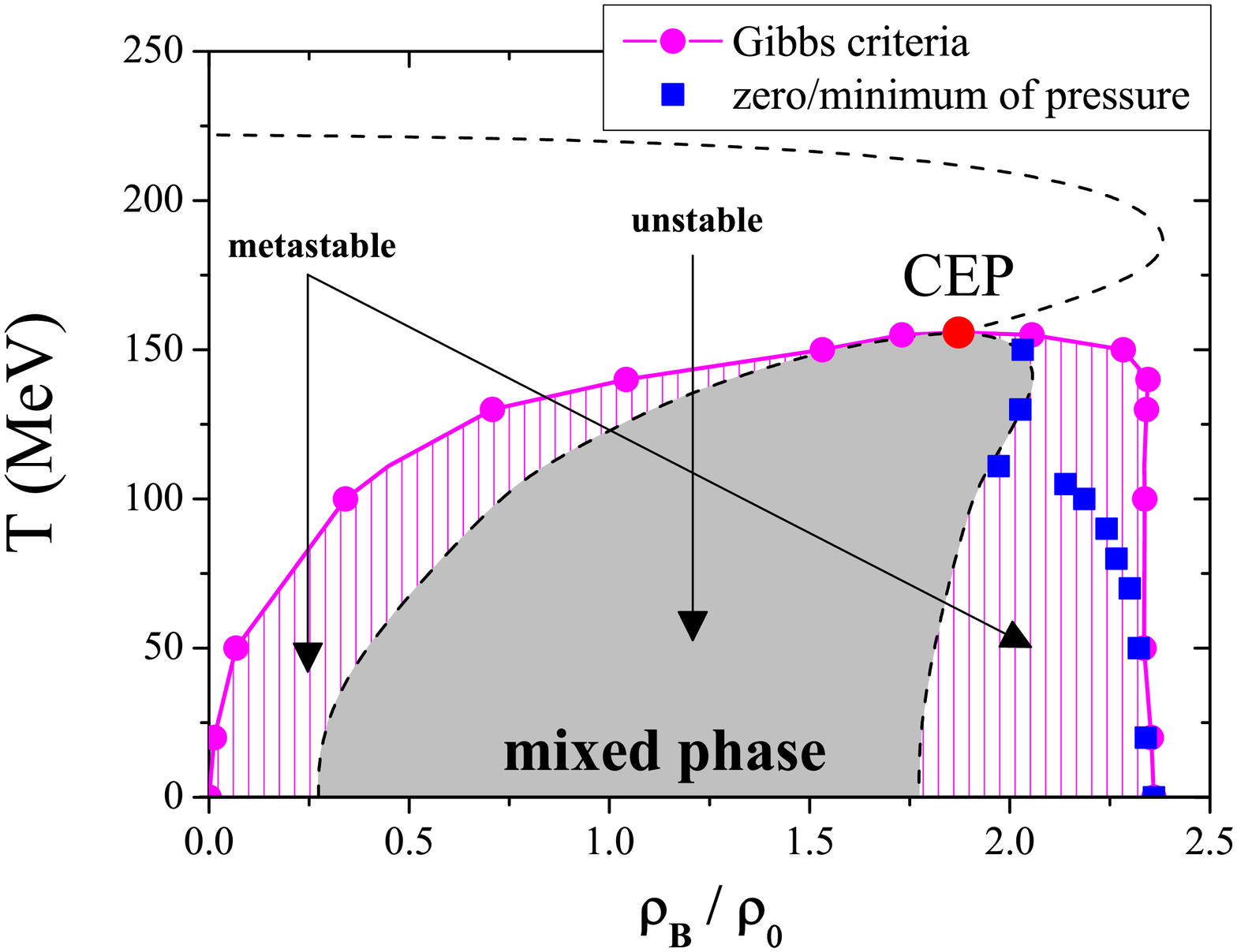,width=9cm,height=7.5cm} \\
   \end{tabular}
\end{center}
\vspace*{-0.5cm}
\caption{
Phase diagram in the SU(3) PNJL model. The left (right) part
corresponds to the $T-\mu_B$ ($T-\rho_B$) plane. Solid (dashed) line shows the location
of the first order (crossover) transition. The dashed lines shows the location of
the spinodal boundaries of the two phase transitions (shown by shading in the right
plot).}
\label{Fig:diag_fase}
\end{figure}

In view of  the behavior  above described, we can conclude that, for
$T=0$, the uniform non-zero density  phase will break up into stable
droplets, with zero pressure and density $\rho_B^{cr} = 2.36
\rho_0$, in which chiral symmetry is partially restored, surrounded
by a nontrivial vacuum with $\rho_B=P=0$ (see also
\cite{Buballa:1998ky,Mishustin:2000ss,Buballa:2003qv,Costa:2003uu,Rajagopal:1999cp,Berges:1998rc}).
In fact, for our choice of the parameters the critical point at
$T=0$ satisfies to the condition $\mu_i<M_i^{vac}$
\cite{Buballa:2003qv,Scavenius:2000qd}, where $M_i^{vac}$ is the
mass of the $i$-quark in the vacuum. This can be seen  by comparing
$\mu_B^{cr}=361.7$ MeV (see the $T$-axis of Figure
\ref{Fig:diag_fase}, left panel) with the quark masses
$M_u^{vac}\,=\,M_d^{vac}\,=\,367.7$ MeV and $M_s^{vac}\,=\,549.5$
MeV.

%%%%%%%%%%%%%%%%%%%     FIG. 6
%%%%%%%%%%%%%%%%%%%

%======================================================================
\subsection{Phase Transition at Finite Temperature and Density/chemical Potential}

Let us now analyze what happens  at $T\neq 0$. As the temperature
increases (see  Figure \ref{Fig:press_energ}, left panel) the first
order transition  persists up to the CEP. At the CEP the chiral
transition becomes a second order one. Along the line of a first
order phase transition the thermodynamic potential has two
degenerate minima, that are separated by a finite potential barrier
making the potential non-convex. The height of the barrier decreases
as the temperature increases and disappears at the CEP. Again, this
pattern is characteristic of a first order phase transition: the two
minima correspond, respectively, to the phases of broken and
restored symmetry. The borders of the coexistence area are marked by
the dotted lines in  Figure \ref{Fig:diag_fase} which shows the $T -
\mu_B$ phase diagram. The domain between the two dashed lines has
metastable states which are characterized by large fluctuations.
They are also solutions of the gap equations but their thermodynamic
potential is higher than for the stable solutions. The left dashed
curves represent the beginning of the metastable solutions of
restored symmetry in the phase of broken symmetry, while the right
dashed curves represent the end of the metastable solutions of
broken symmetry in the restored symmetric phase. We also represent
in Figure \ref{Fig:diag_fase} (right panel) the region where the
solutions of the gap equations are unstable.

The location of the CEP is found to be at $T^{CEP} = 155.80$ MeV and
$\rho_B^{CEP}=1.87\rho_0$ ($\mu_B^{CEP} = 290.67$ MeV).
For temperatures above the CEP the thermodynamic potential has only
one minimum and the transition is washed out: a smooth crossover
takes place. In the left panel of Figure \ref{Fig:diag_fase} the 
crossover is represented by the dashed-dotted curve. 
To calculate the crossover we took the zero of
the ${\partial^2 \left\langle \bar{q_u}q_u \right\rangle}/{\partial
T^2}$, {\em i.e.} the inflection point of the quark condensate
$\left\langle \bar{q_u}q_u \right\rangle$. The crossover is defined
as the rapid decrease of the quark condensate (the order-like
parameter). To determine the range of the crossover region we choose
to define it as the interval between the two peaks around the zero
of ${\partial^2 \left\langle \bar{q_u}q_u \right\rangle}/{\partial
T^2}$. This area is presented in gray in Figure \ref{Fig:diag_fase}
(left panel). We can see that as $\mu_B$ increases the area where
the crossover takes place is narrowed until we reaches the CEP.

Finally, we will focus again on the energy per baryon. In Figure
\ref{Fig:press_energ} (right panel), we plot the density dependence
of the energy per baryon at different temperatures.
We observe that the two points (zero of the pressure and  minimum of
the energy density) are not the same at finite temperature. In fact,
as can be seen from Figure \ref{Fig:press_energ} (left panel),
states with zero pressure are only possible up to the maximal
temperature $T_m \sim 111$ MeV.
For $T<T_m$ the zero-pressure states are in the metastable density
region and, as soon as $T\neq 0$, they do not coincide with the
minimum of the energy per particle (see Figure
\ref{Fig:press_energ}, right panel).

The  arguments just presented allow to clarify the difference
between confined quark matter (in hadrons) and bounded quark matter
(droplets of quarks).

This pattern of phase transition is similar to the liquid--gas
transition in nuclear matter, a consequence of the fact that nuclear
matter assumes its ground state at a non-vanishing baryon density
$\rho_B\sim 0.17$ fm$^{-3}$ when $T=0$.

%======================================================================
%======================================================================

\section{Nernst Principle and Isentropic Trajectories} \label{sec_isent}

The isentropic lines contain important information about the
conditions that are supposed to be realized in heavy ion collisions.
Most of the studies on this topic have been done with lattice
calculations for two flavor QCD at finite $\mu$ \cite{Ejiri:2005uv}
but there are also studies using different type of models
\cite{Scavenius:2000qd,Nonaka:2004pg,Kahara:2008yg}. Some model
calculations predict that in a region around the CEP the properties
of matter are only slowly modified as the collision energy is
changed, as a consequence of the attractor character of the CEP
\cite{Stephanov:1998dy}.

\begin{figure}[t]
    \vspace{-0.5cm}
    \begin{center}
        \includegraphics[width=0.65\textwidth]{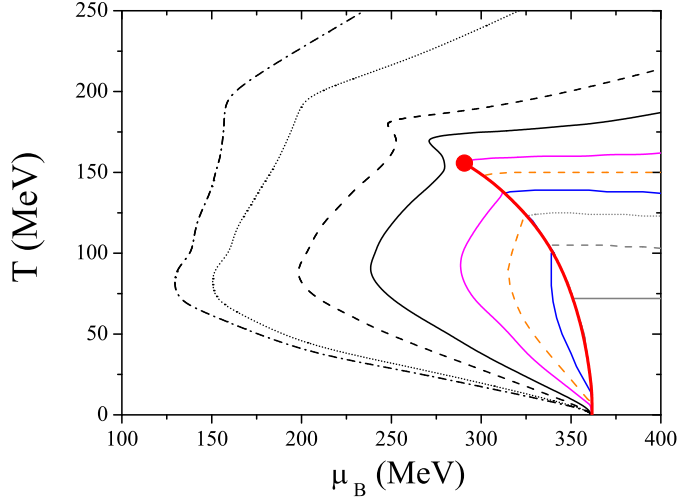}
    \end{center}
    \vspace{-0.5cm}
\caption{Isentropic trajectories in the $(T,\mu_B)$ plane. The following
values of the entropy per baryon number have been  considered:
$s/\rho_B=1,\,2,\,3,\,4,\,5,\,6,\,8,\,10,\,15,\,20$ (anticlockwise direction).}
\label{Fig:isent}
\end{figure}

Our numerical results for the isentropic lines in the $(T,\mu_B)$
plane are shown in Figure \ref{Fig:isent}. We start the discussion
by analyzing the behavior of the isentropic lines in the limit
$T\rightarrow 0$. As already analyzed in Section \ref{phase}, our
convenient choice of the model parameters allows a better
description of the first order transition than other treatments of
the NJL (PNJL) model. This choice is crucial to obtain important
results: the criterion of stability of the quark droplets
\cite{Buballa:2003qv,Costa:2003uu} is fulfilled, and, in addition,
simple thermodynamic expectations in the limit $T\rightarrow 0$ are
verified.
In fact, in this limit $s \rightarrow 0$ according to the third law
of thermodynamics and, as $\rho_B \rightarrow 0$ too, the
satisfaction of the condition $s/\rho_B\,=\,const.$ is insured.
We recall (Section \ref{phase})  that, at $T=0$, we are in the
presence of droplets (states in mechanical equilibrium with the
vacuum state ($\rho_B=0$) at $P=0$).

%%%%%%%%%%%%%%%%%%%     FIG. 7
%%%%%%%%%%%%%%%%%%%

At $T\neq0$, in the first order line, the behavior we find is
somewhat different from those claimed by other authors
\cite{Nonaka:2004pg,Stephanov:1999zu} where a phenomena of focusing
of trajectories towards the CEP is observed. We see that the
isentropic lines with $s/\rho_B=1,...,6$ come from the region of
symmetry partially restored and attain directly the phase
transition; the trajectory  $s/\rho_B=1$ goes along with the phase
transition as $T$ decreases until it reaches $T=0$; and the other
trajectories enter the hadronic phase where the symmetry is still
broken and, after that,  also converge to the horizontal axes
($T=0$).
Consequently, even without reheating in the mixed phase as verified
in the ``zigzag'' shape of
\cite{Ejiri:2005uv,Nonaka:2004pg,Kahara:2008yg,Subramanian:1986xh},
all isentropic trajectories directly terminate in the end of the
first order transition line at $T=0$.

In the crossover region the behavior of the isentropic lines is
qualitatively similar to the one obtained in lattice calculations
\cite{Ejiri:2005uv} or in some models
\cite{Nonaka:2004pg,Fukushima:2009dx,Kahara:2008yg}.
The trajectories with  $s/\rho_B>6$ go directly to the crossover
region and display a smooth behavior, although those that pass in
the neighborhood of the CEP show a slightly kink behavior.

In conclusion,  all the trajectories  directly terminate in the same
point of the horizontal axes at $T=0$.
As already pointed out in \cite{Scavenius:2000qd}, the picture
provided here is a natural result in these type of quark models with
no change in the number of degrees of freedom of the system in the
two phases. As the temperature decreases a first order phase
transition occurs, the latent heat increases and  the formation of
the mixed phase is thermodynamically favored.

We point out again  that, in the limit $T\rightarrow 0$, it is
verified that $s \rightarrow 0$ and $\rho_B \rightarrow 0$, as it
should be.
This behavior is in contrast to the one reported in \cite{Scavenius:2000qd} 
(see right panel of Fig. 9 therein), where the NJL model in the SU(2) sector is used.
The difference is due to our more convenient choice of the model
parameters, mainly a lower  value of the cutoff. This can be
explained by the presence of droplets at $T=0$ whose stability is
quite sensitive to the choice of the model parameters.

%======================================================================
%======================================================================
\section{Effects of Strangeness and Anomaly Strength on the Critical End Point}\label{CEP1}

As already noticed, an important question which emerges in the study
of the QCD phase diagram  is whether the hadronic phase and the
quark gluon like phase are separated by a phase transition in the
thermodynamical sense. The identification of possible critical
points associated to the phase transition is a challenging problem.
Since we are working within a model with (2+1) flavors with the
U$_A$(1) anomaly included, we can explore  new  scenarios with
regards  to the usual SU(2) model without explicit anomaly. We have
now two more parameters: the strange current quark mass, $m_s$, and
the anomaly coupling strength, $g_D$, and it is certainly
interesting to study the effect on thermodynamic quantities of
changing these parameters. The anomaly is present via the 't Hooft
interaction and its effects appear explicitly in the gap equations
(\ref{eq:gap}) and in the expression of the thermodynamical
potential (\ref{omega}) through products of the anomaly coefficient
by quark condensates. Here we will discuss the influence on the
location and the nature of the critical points of the degree of
explicit symmetry breaking of both chiral symmetry (in the strange
sector) and axial U$_A$(1) symmetry.

%======================================================================
\subsection{Role of the Strangeness in the Location of the CEP/TCP}

Let us study the influence of the strangeness in the location of the
CEP/TCP. As we have already  seen, the location of the CEP is found
at $T^{CEP}=155.80$ MeV and $\mu_B^{CEP} = 290.67$ MeV
($\rho_B^{CEP}=1.87\rho_0$) when we use physical values of the quark
masses \cite{Rehberg:1995kh,Costa:2007ie}: $m_u = m_d = 5.5$ MeV,
$m_s = 140.7$ MeV.

\begin{figure}[t]
\begin{center}
\vspace{-0.5cm}
  \begin{tabular}{cc}
    \epsfig{file=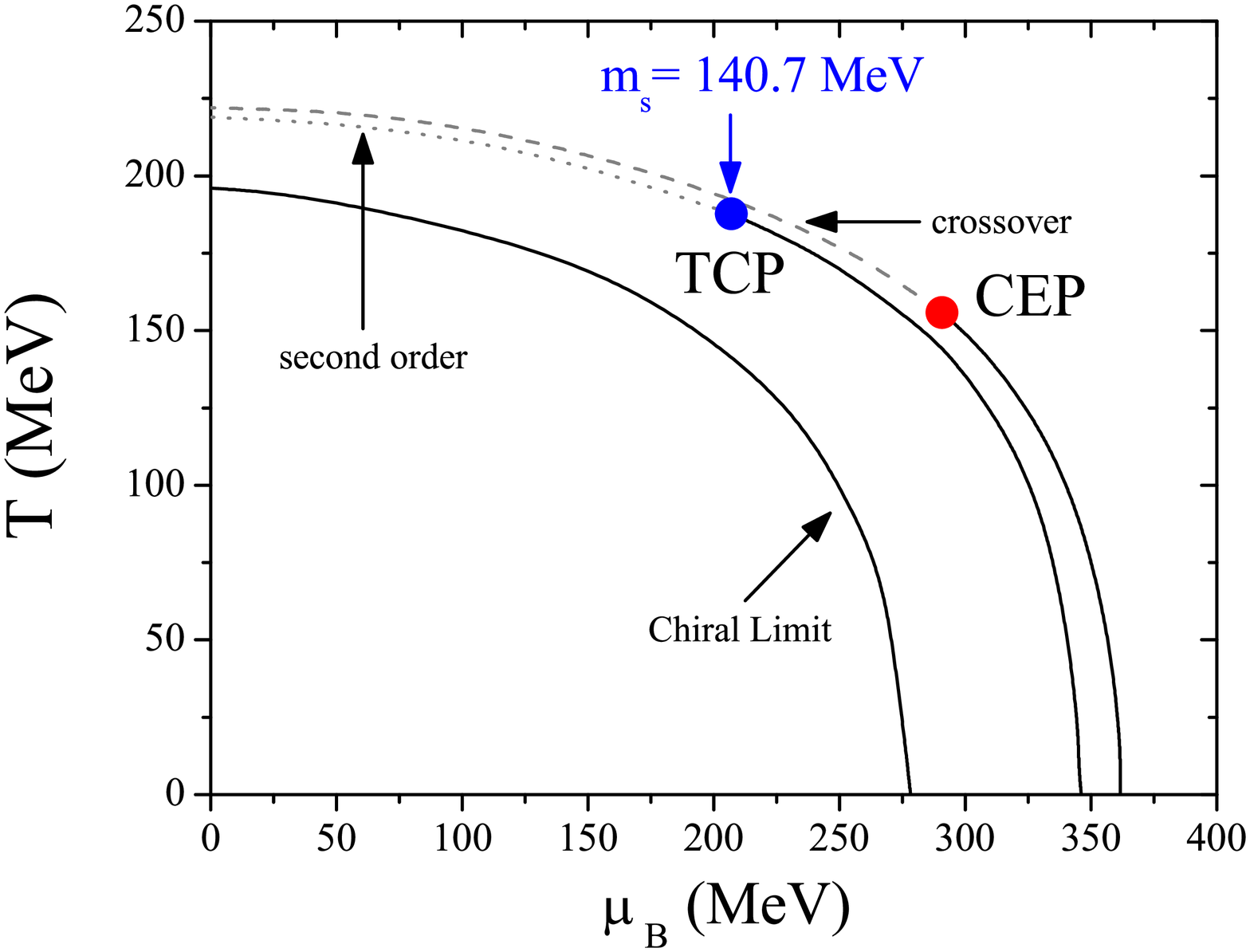,width=9cm,height=7.5cm} &
    \hspace*{-1.0cm}\epsfig{file=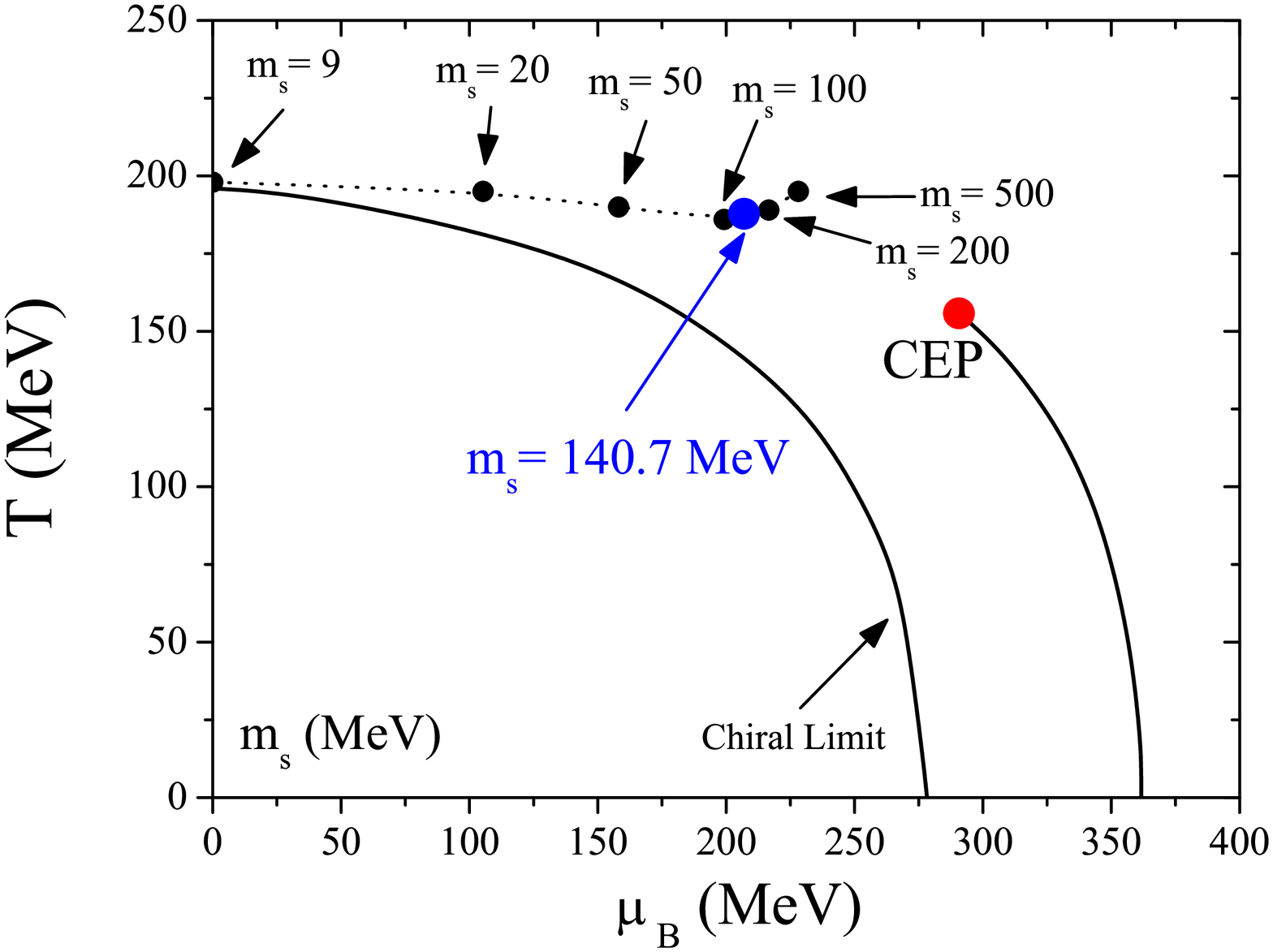,width=9cm,height=7.5cm} \\
   \end{tabular}
   \vspace{-0.5cm}
\end{center}
\caption{Left panel: the phase diagram
in the SU(3) PNJL model. The solid lines represent the first order
phase transition, the dotted line the second order phase transition,
and the dashed line the crossover transition. Right panel: the phase
diagram and the ``line'' of TCPs for  different values of $m_s$ (the
dotted lines are just drawn to guide the eye); the  TCPs in both
figures are obtained  in the limit $m_u=m_d=0$ and $m_s\neq 0$.}
\label{Fig:diagfases}
\end{figure}

A first point to be noticed is that in the PNJL model, contrarily to
what happens in the chiral limit only for the SU(2) sector
($m_u=m_d=0$, $m_s\neq 0$) where the TCP is found
\cite{Costa:2009ae}, when the total chiral limit is considered
($m_u=m_d=m_s=0$), the phase diagram does not exhibit a TCP: chiral
symmetry is restored via a first order transition for all baryonic
chemical potentials and temperatures (see left panel of Figure
\ref{Fig:diagfases}). Both situations are in agreement with what is
expected: the chiral phase transition at the chiral limit is of
second order for $N_f = 2$ and first order for $N_f\geq3$
\cite{Pisarski:1983ms}.

To study the influence of strangeness on the location of the
critical points, we vary the current quark mass $m_s$, keeping the
SU(2) sector in the chiral limit and the other model parameters
fixed. The phase diagram is presented in Figure \ref{Fig:diagfases}
(right panel) as a function of $\mu_B$ and $T$, and we consider
different cases for the current quark mass $m_s$.

%%%%%%%%%%%%%%%%%%%     FIG. 8
%%%%%%%%%%%%%%%%%%%

The pattern of chiral symmetry restoration via first order phase
transition remains for $m_u=m_d=0$ and $m_s<m_{s}^{crit}$
\cite{Hsu:1998eu}. The value  for $m_s^{crit}$ is a subject of
debate; those found in lattice \cite{Laermann:2003cv} or in model
calculations \cite{Hsu:1998eu,Barducci:2005ut} are lower than the
physical strange current quark mass ($m_s\approx 150$ MeV). We found
$m_s^{crit}\approx 9$ MeV in our model, lower than lattice values
\cite{Laermann:2003cv} and half of the value obtained in NJL model
($m_s^{crit} = 18.3$ MeV \cite{Costa:2007ie}), but still consistent
with other models of this type \cite{Barducci:2005ut}.
When $m_s\geq m_{s}^{crit}$, at $\mu_B=0$, the transition is of the
second order and, as $\mu_B$ increases, the line of the second
order phase transition will end in a first order line at the TCP.
Several TCPs are plotted  for different values of $m_s$ in the right
panel of Figure \ref{Fig:diagfases}. As $m_s$ increases, the value
of $T$ for this ``line'' of TCPs decreases as $\mu_B$ increases
getting closer to the CEP and, when $m_{s}=140.7$ MeV, it starts to
move away from the CEP. The TCP for $m_{s}=140.7$ MeV is the closest
to the CEP and is located at $\mu_B^{TCP}=206.95$ MeV and
$T^{TCP}=187.83$ MeV. If we choose $m_u=m_d\neq0$, instead of a
second order transition we have a smooth crossover for all the
values of $m_s$ and the ``line'' of TCPs becomes a ``line'' of CEPs.

%======================================================================
\subsection{Role of the Anomaly Strength in the Location of the CEP}

The axial U$_A$(1) symmetry is broken explicitly by instantons,
leaving a SU(N$_f)\times$ SU(N$_f)$ symmetry which determines the
chiral dynamics. Since instantons are screened in a hot or dense
environment, the U$_A$(1) symmetry may be effectively restored in
matter. So, it is instructive to review this enlargement of chiral
symmetry to SU(N$_f)\times$ SU(N$_f)\times$U$_A$(1) in more detail.

\begin{figure}[t]
\begin{center}
\vspace{-0.5cm}
    \includegraphics[width=0.65\textwidth]{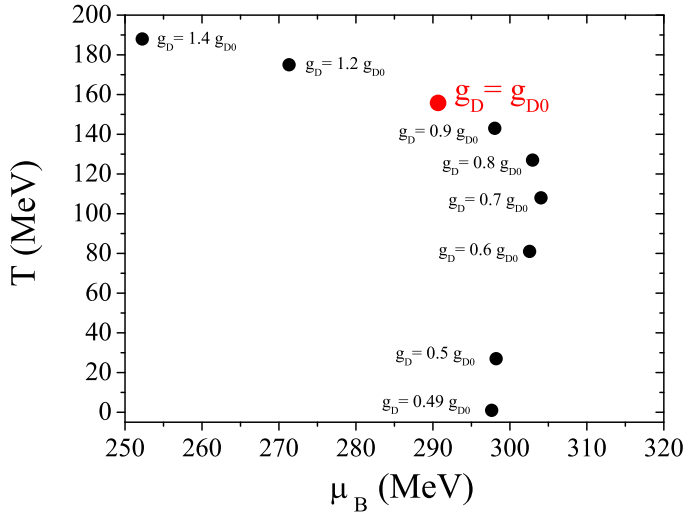}
\vspace{-0.5cm}
\end{center}
\caption{Dependence of the location of the CEP on the strength of
the 't Hooft coupling constant $g_D$.}
\label{Fig:CEP_GD}
\end{figure}

As already referred,  the location and even the existence of the CEP
in the phase diagram is a matter of debate \cite{CEP}. While
different lattice calculations predict the existence of a CEP
\cite{Fodor:2004nz}, the absence of the CEP in the phase diagram was
seen in recent lattice QCD results
\cite{deForcrand:2006pv,deForcrand:2007rq,deForcrand:2008vr}, where
the first order phase transition region near $\mu_B=0$ shrinks in  the
quark mass and $\mu_B$ space when $\mu_B$ is increased
\cite{deForcrand:2006pv,deForcrand:2007rq,deForcrand:2008vr}. Due to
the importance of the  U$_A$(1) anomaly, already emphasized in
Section \ref{QCD},  and its influence on several observables, it is
demanding to investigate possible changes in the location of the CEP
in the $(T,\,\mu_B)$ plane when the anomaly strength is modified. In
Figure \ref{Fig:CEP_GD} we show the location of the CEP for several
values of $g_D$ compared to the results for $g_{D_0}$, the value
used for the vacuum. As already pointed out by K. Fukushima in
\cite{Fukushima:2008wg}, we also observe that the location of the
CEP depends on the value of $g_D$ and, when $g_D$ is about 50$\%$ of
its value in the vacuum, the QCD critical point disappears from the
phase diagram; the first order region becomes narrower with the
decreasing of the $g_D$ value.

%%%%%%%%%%%%%%%%%%%     FIG. 9
%%%%%%%%%%%%%%%%%%%

The results show that, in the framework of this model, the existence
or not of the CEP is determined by the strength of the anomaly
coupling, the CEP getting closer to the $\mu_B$ axis as $g_D$
decreases.

%======================================================================
%======================================================================
\section{Susceptibilities and Critical Behavior in the Vicinity of the CEP}\label{CEP2}

In the last years, the phenomenological relevance of fluctuations in
the finite temperature and chemical potential around the CEP/TCP of
QCD has been attracting the attention of several authors
\cite{Hatta:2002sj}. As a matter of fact, fluctuations are supposed
to represent signatures of phase transitions of strongly interacting
matter. In particular, the quark number susceptibility plays a role
in the calculation of event-by-event fluctuations of conserved
quantities such  as the net baryon number. Across the quark hadron
phase transition they are expected to become large; that can be
interpreted as an indication for a critical behavior. We also
remember the important role of the second derivative of the pressure
for second order points like the CEP.

The grand canonical potential (or the pressure) contains the
relevant information on thermodynamic bulk properties of a medium.
Susceptibilities, being second order derivatives of the pressure in
both chemical potential and temperature  directions, are related to
those fluctuations. The relevance of these physical observables is
related with the size of the critical region around the CEP which
can be found by calculating the specific heat, the baryon number
susceptibility, and their critical behaviors. The size of this
critical region is important for future searches of the CEP in heavy
ion-collisions \cite{{Nonaka:2004pg}}.

\begin{figure}[t]
    \begin{center}
    \vspace{-0.5cm}
        \includegraphics[width=0.65\textwidth]{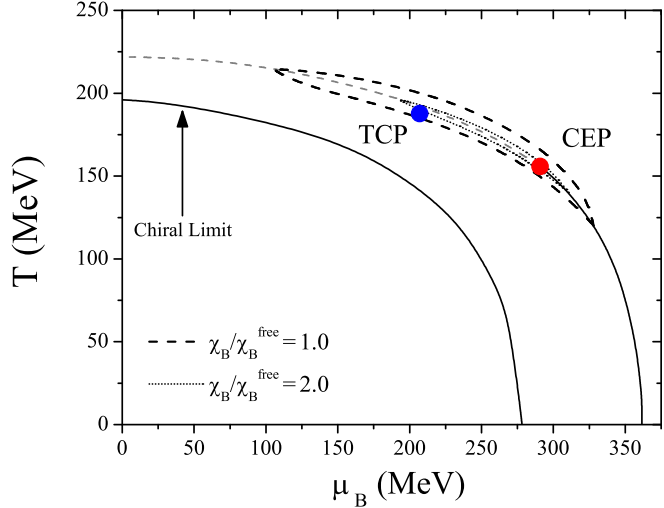}
    \end{center}
    \vspace{-0.5cm}
\caption{Phase diagram: the size of the critical region is plotted for $\chi_B/\chi_B^{free}=1(2)$. 
The TCP is found for $m_u=m_d=0$ MeV and $m_{s}=140.7$ MeV.}
\label{Fig:zona_critica}
\end{figure}

The way to estimate the critical region around the CEP is to
calculate the dimensionless ratio $\chi_B/\chi_B^{free}$, where
$\chi_B^{free}$ is the chiral susceptibility of a free massless
quark gas.
Figure \ref{Fig:zona_critica} shows a contour plot for two fixed
ratios $\chi_B/\chi_B^{free}=1.0$ and $2.0$ in the phase diagram
around the CEP.
In the direction parallel to the first order transition line and to
the crossover, it can be seen an elongation of the region where
$\chi_B$ is enhanced, indicating that the critical region is heavily
stretched in that direction.
It means that the divergence of the correlation length at the CEP
affects the phase diagram quite far from the CEP and a careful
analysis including effects beyond the mean-field needs to be done
\cite{Rossner:2007ik}.

%%%%%%%%%%%%%%%%%%%     FIG. 10
%%%%%%%%%%%%%%%%%%%

One of the main effects of the Polyakov loop is to shorten the
temperature range where the crossover occurs \cite{Hansen:2006ee}.
On the other hand, this behavior is boosted by the choice of the
regularization ($\Lambda \rightarrow \infty$) \cite{Costa:2009ae}.
The combination of both effects results in higher baryonic
susceptibilities even far from the CEP when compared with the NJL
model \cite{Costa:2008gr}.
This effect of the Polyakov loop is driven by the fact that the one-
and two-quark Boltzmann factors are controlled by a factor
proportional to $\Phi$: at small temperature $\Phi \simeq 0$ results
in a suppression of these contributions (see Equation (\ref{omega}))
leading to a partial restoration of the color symmetry. Indeed, the
fact that only the $3-$quark Boltzmann factors $e^{3\beta E_p}$
contribute to the thermodynamical potential at low temperature, may
be interpreted as the production of a thermal bath containing only
colorless 3-quark contributions.  When the temperature increases,
$\Phi$ goes quickly to 1 resulting in a (partial) restoration of the
chiral symmetry occurring in a shorter temperature range.
The crossover taking place in a smaller $T$ range can be interpreted
as a crossover transition closest to a second order one. This
``faster'' crossover may explain the elongation of the critical
region giving raise to a greater correlation length even far from
the CEP.

\begin{figure}[t]
\begin{center}
\vspace{-0.5cm}
  \begin{tabular}{cc}
    \epsfig{file=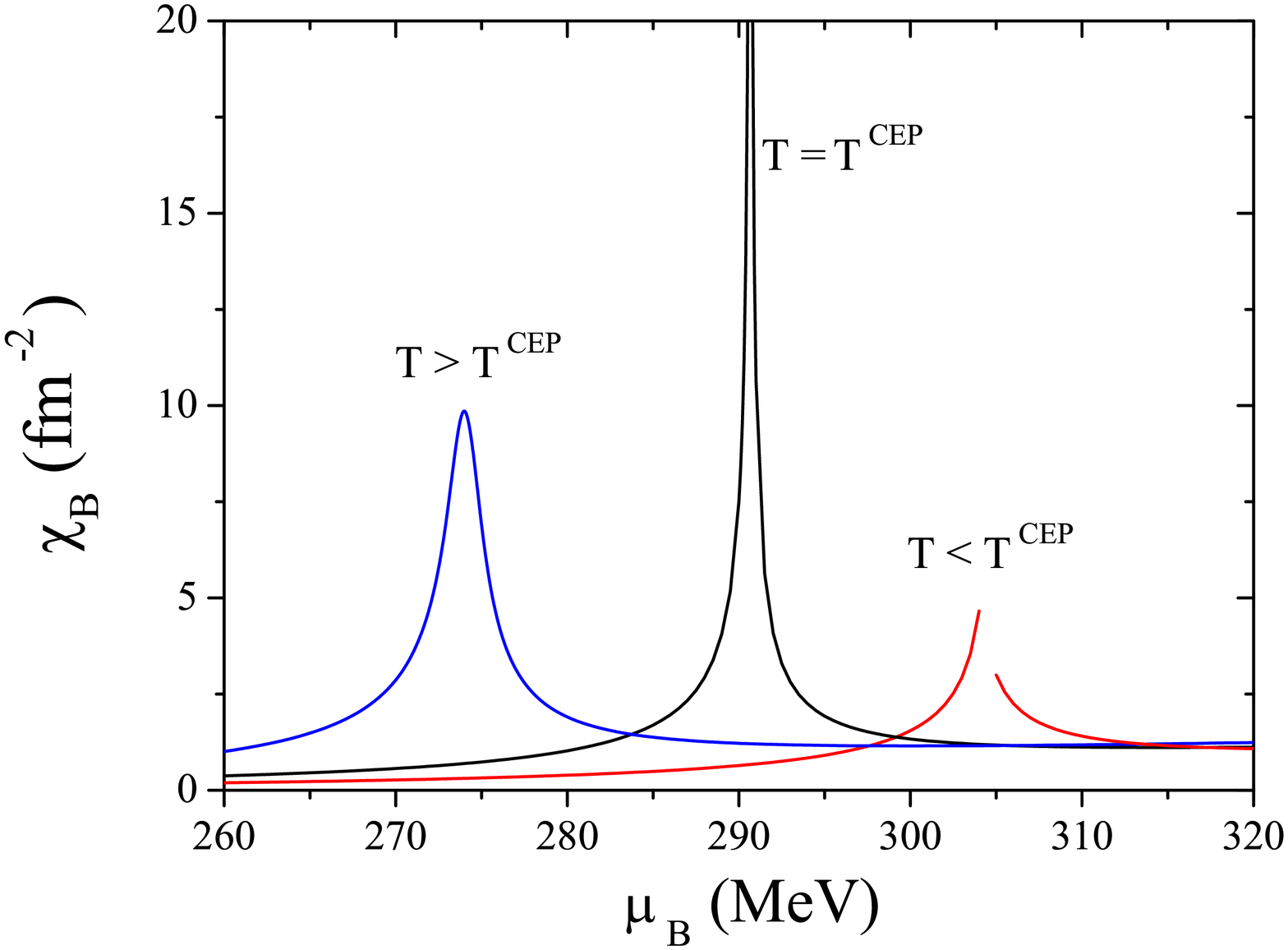,width=9cm,height=7.5cm} &
    \hspace*{-1cm}\epsfig{file=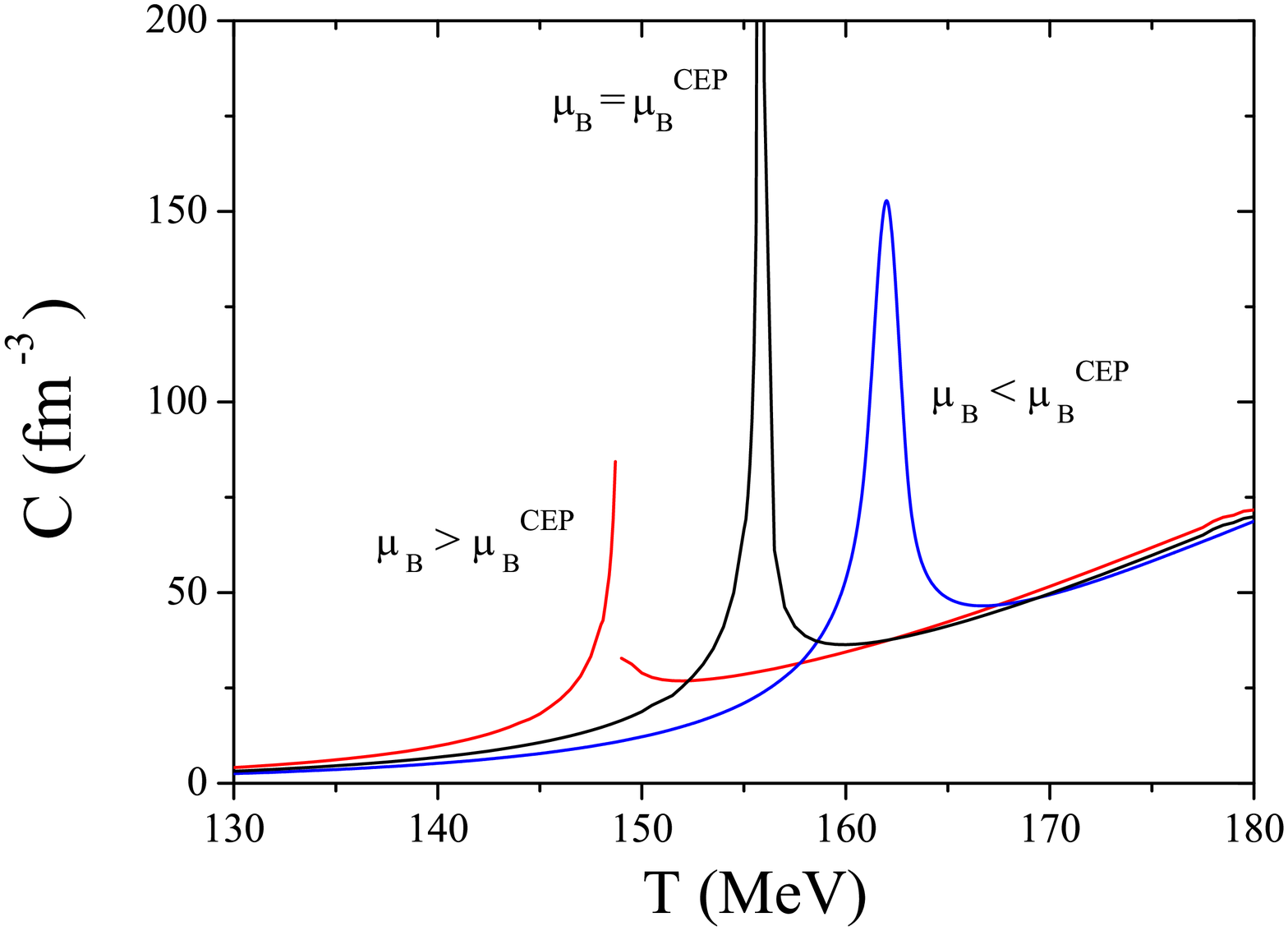,width=9cm,height=7.5cm} \\
   \end{tabular}
   \vspace{-0.5cm}
\end{center}
\caption{Left panel: Baryon number susceptibility (right panel)
         as functions of $\mu_B$ for different temperatures around the CEP: $T^{CEP} = 155.80$ MeV
         and $T=T^{CEP}\pm10$ MeV. Right panel: Specific heat as a function of $T$ for different
         values of $\mu_B$ around the CEP: $\mu_B^{CEP} = 290.67$ MeV and $\mu_B=\mu_B^{CEP}\pm10$ MeV.}
\label{Fig:chi_calor}
\end{figure}

Now we show  in Figure \ref{Fig:chi_calor} the  behavior of the
baryon number susceptibility, $\chi_B$, and the specific heat, $C$,
at the CEP,  which is in accordance with
\cite{Hatta:2002sj,Schaefer:2006ds,Costa:2007ie}.

%%%%%%%%%%%%%%%%%%%     FIG. 11
%%%%%%%%%%%%%%%%%%%

The baryon number susceptibility is plotted (left panel) as a
function of the baryon chemical potential for three values of the
temperature. The divergent behavior of $\chi_B$ at the CEP is an
indication of the second order phase transition at this point. The
curve for $T>T^{CEP}$ corresponds to the crossover and the other to
the first order transition.

Now we will pay attention to the specific heat (Equation
(\ref{chi})) which is plotted as a function of the temperature
around the CEP (right panel). The divergent behavior is the signal
of the location of the CEP. The curve $\mu_B < \mu_B^{CEP}$
corresponds to the crossover and the other to the first order
transition.

The large enhancement of the baryon number susceptibility and the
specific heat at the CEP may be used as a signal of the existence
and identification of phase transitions in the quark matter.

%======================================================================
%======================================================================
\section{Conclusions}\label{conclusions}

The concept of symmetries is a very important topic in physics. It
has given a fruitful insight into the relationships  between
different areas, and  has contributed to the unification of several
phenomena, as shown in recent achievements in nuclear and hadronic
physics.

Critical phenomena in hot QCD have  been studied in the framework of
the PNJL model, as an important issue to determine the order of the
chiral phase transition as a function of the temperature  and the
quark chemical potential.
Symmetry arguments show that the phase transition should be a first
order one  in the chiral limit $(m_u=m_d = m_s = 0)$. Working out of
the chiral limit,   at which both chiral and center symmetries are
explicitly broken, a CEP which separates  first and crossover lines
is found, and the corresponding order parameters are analyzed.

The sets of parameters  used  is compatible with the formation of
stable droplets at zero temperature, insuring  the satisfaction of
important thermodynamic expectations like the Nernst principle. Other
important role is played by the regularization procedure which, by
allowing high momentum quark states, is essential to obtain the
required increase of extensive thermodynamic quantities, insuring
the convergence to the Stefan--Boltzmann (SB) limit of QCD. In this
context the  gluonic degrees of freedom also play a special role.

We also discussed the effect of the U$_A$(1) axial symmetry both at
zero and at finite temperature. We analyzed the effect of the
anomalous coupling strength in the location of the CEP. We proved
that the location of the CEP depends on the value of $g_D$ and, when
$g_D$ is about 50$\%$ of its value in the vacuum, the QCD critical
point disappears from the phase diagram.
One expects that, above a certain critical temperature $T_{eff}$,
the chiral and axial symmetries will be effectively restored. The
behavior of some given observables signals the effective restoration
of these symmetries: for instance, the topological susceptibility
vanishes.

The successful comparison with lattice results shows that the model
calculation provides a convenient tool to obtain information for
systems from zero to non-zero chemical potential which is of
particular importance for the knowledge of the equation of state of
hot and dense matter. Although the results here presented relies on
the chiral/deconfinement phase transition, the relevant physics
involved is also  useful to understand other phase transitions
sharing similar features.

%%%%%%%%%%%%%%%%%%%%%%%%%%%%%%%%%%%%%%%%%%%%%%%%%%%%%%%%%%%%
\section*{Acknowledgements}
Work supported by  Centro de F\'{\i}sica Computacional and F.C.T. 
under Project No. CERN/FP/83644/2008.

%==========================================================
\section*{Appendix}\label{appendix}

The quark condensates $\left\langle\bar{q_{i}}q_{i}\right\rangle$, with $i,j,k=u,d,s$ (to
be fixed in cyclic order),  determined in a self-consistent way, are given by:
%%%%%%
\begin{equation}
\left\langle\bar{q}_{i}q_{i}\right\rangle\,=\,-\,\,2N_c\int\frac{\mathrm{d}^3p}{\left(2\pi\right)^3}
\frac{M_i}{E_i}[\theta(\Lambda^2-\vec{p}^{2})-f^{(+)}_\Phi(E_i)-f^{(-)}_\Phi(E_i)]
\end{equation}
where $E_i$ is the quasi-particle energy for the quark $i$:
$E_{i}=\sqrt{\mathbf{p}^{2}+M_{i}^{2}}$.

%%%%%%
All calculations in the NJL model can be generalized to the PNJL one, as was shown in
\cite{Hansen:2006ee}, by introducing the modified Fermi-Dirac distribution functions for
particles and antiparticles used in the expression of
$\left\langle\bar{q}_{i}q_{i}\right\rangle$:
%%%%%%
\be f^{(+)}_\Phi(E_i) & = &\frac{ \bar\Phi\expp + 2\Phi\exppp + \expppp }
  {\exp\{z^+_\Phi(E_i)\}} \\
f^{(-)}_\Phi(E_i) & = &\frac{ \Phi\expm + 2\bar\Phi\expmm + \expmmm }
  {\exp\{z^-_\Phi(E_i)\}} \label{fpPhi}
\ee

The  mean field equations are obtained by minimizing the  thermodynamical potential
density (\ref{omega}) with respect to $\left\langle\bar{q}_{i}q_{i}\right\rangle$
($i=u,d,s$), $\Phi$, and $\bar\Phi$. The respective mean field equations are the already
given gap equation (\ref{eq:gap}) and

%%%%%%
\be 
0 &=& {T^4} \left\{ -\frac{a(T)}{2}\bar\Phi - 6\frac{b(T)\left[\bar\Phi
-2\Phi^2+\bar\Phi^2 \Phi\right]}
{1-6 \bar\Phi \Phi + 4(\bar\Phi^3 + \Phi^3)-3(\bar\Phi \Phi)^2}\right\} \nonumber\\
 &-& 6 T \sum_{\left\{i=u,d,s\right\}}
   \int\frac{\mathrm{d}^3p}{\left(2\pi\right)^3}
   \left( \frac {\exppp}{ \exp\{z^+_\Phi(E_i)\} } + \frac {\expm}{\exp\{ z^-_\Phi(E_i)\} } \right)
\label{eq:domegadfi} 
\ee
%%%%%%
%%%%%%
\be 
0 &=& {T^4} \left\{ -\frac{a(T)}{2}\Phi - 6\frac{b(T)\left[\Phi -2\bar\Phi^2+\bar\Phi
\Phi^2\right]} {1-6 \bar\Phi \Phi + 4(\bar\Phi^3 + \Phi^3)-3(\bar\Phi \Phi)^2}\right\}
\nonumber\\
   &-& 6 T \sum_{\left\{i=u,d,s\right\}}
   \int\frac{\mathrm{d}^3p}{\left(2\pi\right)^3}
   \left( \frac {\expp}{ \exp\{z^+_\Phi(E_i)\} } + \frac {\expmm}{\exp\{ z^-_\Phi(E_i)\} } \right)
\label{eq:domegadfib} 
\ee
%%%%%%

%%%%%%%%%%%%%%%%%%%%%%%%%%%%%%%%%%%%%%%%%%%%%%%%%%%%%%%%%%%%
%%%%%%%%%%%%%%%%%%%%%%%%%%%%%%%%%%%%%%%%%%%%%%%%%%%%%%%%%%%%
%\bibliography{biblio}        %or whatever your .bib file is
%%\bibliographystyle{h-physrev3}   
%\bibliographystyle{apsrev}

\end{document}